# Symmetry-broken superconducting configurations from density functional theory for bcc and hcp metals and Nb$_3$Sn

Shun-Li Shang and Zi-Kui Liu

Department of Materials Science and Engineering, The Pennsylvania State University,
University Park, PA 16802, USA

**Abstract:**

We recently proposed a unified theoretical framework for superconductivity that broadens the applicability of Bardeen-Cooper-Schrieffer (BCS) theory to both conventional and unconventional superconductors. Within this framework, superconductivity arises from the formation of a *symmetry-broken* superconducting configuration (SCC) generated by atomic perturbations of the normal conducting configuration (NCC). The SCC emerges through electron-phonon interaction and gives rise to distinct straight one-dimensional tunnels (SODTs) in the charge density difference of electrons and/or holes. These SODTs originate from regular and systematic atomic displacements between the SCC and NCC, a phenomenon revealed by density functional theory (DFT) calculations. To further verify this framework, we performed DFT-based calculations for 12 hexagonal close-packed (hcp) elements (Be, Mg, Sc, Y, Ti, Zr, Hf, Tc, Re, Ru, Os, and Zn), 5 body-centered cubic (bcc) elements (V, Nb, Ta, Mo, and W), and the compound Nb$_3$Sn. Our results indicate that all these materials exhibit superconductivity at 0 K and 0 GPa, as evidenced by the predicted SODTs. Notably, Mg, Sc, and Y are predicted to be superconducting under ambient pressure, a finding that awaits experimental confirmation.

**Keywords:**

Bardeen-Cooper-Schrieffer (BCS) theory; Density functional theory (DFT); Superconducting configuration (SCC); Normal conducting configuration (NCC); Straight one-dimensional tunnels (SODTs); Pure element superconductors; Nb$_3$Sn

## 1 Introduction

Superconductivity was first discovered in 1911 by Heike Kamerlingh Onnes in mercury (Hg) at approximately 4.2 K [1]. Nearly half a century later, in 1957, Bardeen, Cooper, and Schrieffer (BCS) [2] developed the microscopic BCS theory to explain superconductivity as a quantum mechanical phenomenon in which electrons form pairs, known as Cooper pairs, below the material's critical temperature $T_C$. These pairs arise through electron-phonon interactions, mediated by lattice vibrations (phonons), enabling electrons to move without electrical resistance. However, superconductivity is inherently fragile due to the weak pairing interaction (on the order of a few meV); consequently, thermal energy can easily break Cooper pairs, limiting conventional superconductors to low $T_C$.

Building on BCS theory, McMillan proposed an approximate formula for the superconducting transition temperature ($T_C$) [3,4],

$$T_C = 0.85 \Theta_D e^{-1/n(\varepsilon_F)\phi_{el-ph}}, \qquad Eq.\ 1$$

where $\Theta_D$ is the Debye temperature related to the highest vibrational frequency, $n(\varepsilon_F)$ is the electronic density of states (eDOS) at the Fermi level, and $\phi_{el-ph}$ is the effective electron-phonon interaction [5,6]. According to Eq. 1, a superconductor must be a conductor with a nonzero $n(\varepsilon_F)$. McMillan further concluded that $T_C$ for conventional superconductors cannot exceed ~ 40 K under ambient pressure, since excessively increasing $\phi_{el-ph}$ destabilizes the lattice [3,7]. Recently, Liu and Shang [8] showed that BCS theory applies to both conventional and unconventional superconductors, noting that electron-phonon coupling may be localized near superconducting layers or directions. Their analysis of the layered "*pontoon*" structure in the high $T_C$ superconductor YBa$_2$Cu$_3$O$_7$ (YBCO7) suggests that weak bonding of superconducting layers to the bulk matrix provides thermal insulation, protecting Cooper pairs from disruption to higher temperatures.

The superconducting gap, representing the energy gain when two electrons form a Cooper pair, is related to $T_C$ for conventional superconductors at 0 K and vanishes at $T_C$ [9,10],

$$\Delta(T = 0K) = 1.764 k_B T_C. \qquad Eq.\ 2$$

This gap is directly linked to the energy difference between the superconducting configuration (SCC) and the normal conducting configuration (NCC), commonly called the condensation energy [2,11,12],

$$\Delta E_{SN} = -\frac{1}{2}n(\varepsilon_F)\Delta^2. \qquad Eq.\ 3$$

Here, $\Delta E_{SN} < 0$ indicates that SCC is the ground state, while $\Delta E_{SN} > 0$ corresponds to a non-superconducting system.

A major theoretical advance after BCS theory was density functional theory (DFT), introduced by Hohenberg and Kohn [13] and Kohn and Sham [14]. DFT provides a practical solution to the many-body Schrödinger equation, enabling prediction of electron-phonon interaction ($\phi_{el-ph}$) in Eq. 1 [15]. DFT asserts that any system has a unique ground-state electron density that minimizing its energy at 0 K [13]. Central to DFT is the exchange-correlation (X-C) functional, which is approximated using schemes such as the local density approximation (LDA), the generalized gradient approximation (GGA), and the meta-GGA [16].

Notably, Sun, Ruzsinszky, and Perdew [17] developed the first nonempirical meta-GGA of the Strongly Constrained and Appropriately Normed (SCAN) functional, satisfying all 17 known exact constraints. In SCAN, strong correlations within a symmetry-unbroken ground-state wavefunction can manifest as symmetry-broken spin or charge densities due to soft fluctuation modes. This concept of independent symmetry-broken configurations was interestingly envisioned by Gibbs in 1902 in developing statistical mechanics [18,19], who "imagined a great number of systems of the same nature, but differing in the configurations and velocities which they have at a given instant, and differing not merely infinitesimally, but it may be so as to embrace every conceivable combination of configuration and velocities".

The main challenge in applying DFT to superconductors lies in identifying the SCC. Inspired by the ground-state electron density principle and the symmetry-breaking concept in SCAN [17] and its regularized version of r$^2$SCAN [20], Liu and Shang [8] proposed a unified theory for both conventional and unconventional superconductors. Their work indicates that the SCC represents a symmetry-broken configuration, and DFT-predicted charge density changes can identify superconductivity at 0 K. Furthermore, $T_C$ can be estimated using the zentorpy theory [21,22], which considers competition between the SCC and NCC. Using these ideas, 14 conventional superconductors were identified among 18 elements with A1, A4, A6, and A7 structures (Al, Pb, Cu, Ag, Au, Rh, Ir, Pd, Pt, *Ca*, *Sr*, *Si*, *Ge*, Sn, In, *As*, Sb, and Bi; where underlined *italic* elements are non-superconducting) and $MgB_2$ at 0 K and 0 GPa; along with one unconventional superconductor of $YBCO_7$ [8]. It is noted that in their work, they missed the more recent experimental observation that Bi is a superconductor with $T_C = 0.53$ mK at ambient pressure [22], which was pointed in private communications by Ramakrishnan, one of the authors of the publication. In those communications, Ramakrishnan also mentioned that they were not able to detect the superconductivity of Sb down to 0.1 mK and Au down to 0.05 mK, probably due to the slightly inferior Sb crystals with residual resistivity ratio of 200 as compared to above 600 for Bi and a few parts per million of Mn in Au.

The present work extends this verification by examining additional conventional superconductors at 0 K and 0 GPa. These include 12 hexagonal close-packed (hcp) elements (Be, Mg, Sc, Y, Ti, Zr, Hf, Tc, Re, Ru, Os, and Zn), 5 body-centered cubic (bcc) elements (V, Nb, Ta, Mo, and W), and the compound $Nb_3Sn$. These elements were chosen for their structural simplicity and available experimental data [23,24] (see Table 1), while $Nb_3Sn$ was included because both its SCC and NCC have been experimentally characterized (Sec. 3.3). For each material, SCC and NCC were generated (Sec. 2.1), DFT-based calculations were performed (Sec. 2.2), and superconductivity was assessed based on predicted straight one-dimensional tunnels (SODTs). All examined materials exhibit superconductivity at 0 K and 0 GPa, including Mg, Sc, and Y, which remain experimentally unconfirmed under ambient conditions (see Sec. 3 and Table 1).

## 2  Methodology

### 2.1  Definition of superconducting configuration (SCC) at 0 K and 0 GPa

Our framework for examining superconductivity using DFT is inspired by several established concepts [8]: Cooper pairing via electron–phonon interactions in the BCS theory; the existence of a ground-state configuration at 0 K and 0 GPa with a unique electron density as described by DFT; and the symmetry-broken configurations introduced by Gibbs [18,19] and later incorporated into meta-GGA functionals such as SCAN [17] and r$^2$SCAN [20].

We define the superconducting configuration (SCC) as a symmetry-broken ground-state at 0 K and 0 GPa, formed through atomic perturbations of the normal conducting configuration (NCC), which dominates above $T_C$ and exhibits higher symmetry. Both SCC and NCC charge densities, which represent the spatial distribution of electrons, can be predicted by DFT. We further propose that the SCC versus NCC charge density difference (SNCDD) form SODTs, acting as "superhighways" for electricity transportation without resistance, analogous to Cooper pairs transport in BCS theory. Electron-phonon interactions in this framework are embedded within DFT calculations of the SCC and NCC, represented by the applied atomic perturbations in SCC.

For practical implementation, materials studied via DFT can be grouped into two categories based on symmetry, which is experimentally measurable using techniques such as X-ray or neutron diffraction:
- Category-1: Higher-symmetry structures with fixed atomic positions (e.g., pure elements with bcc, fcc, and hcp structures, and $Ni_3Sn$ with its A15 cubic structure [25,26]).
- Category-2: Lower-symmetry structures with variable atomic positions (e.g., high-$T_C$ superconductors such as $YBa_2Cu_3O_{7-\delta}$ [27] and $La_{2-x}Ba_xCuO_4$ [28]).

For Category-1 materials, experimentally measured structures are treated as NCC. SCC is generated by randomly perturbing atomic positions [8],

$$(x_0 + n_x\delta_{ini},\quad y_0 + n_y\delta_{ini},\quad z_0 + n_z\delta_{ini}) \qquad Eq.\ 4$$

where $x_0, y_0, z_0$ are atomic coordinates in the NCC; $n_x, n_y$, and $n_z$ are random numbers 0, 1, or -1; and $\delta_{ini}$ typically ranges from 0.02 to 0.08 (fractional coordinates) to prevent atoms from reverting to their original high-symmetry positions. These perturbations aim to mimic strong electron-phonon interactions to identify stable SCCs and typically exceed those used in harmonic phonon and electron-phonon coupling calculations [29,30]. After testing multiple $\delta_{ini}$ values, the configuration yielding the lowest total energy is selected as the SCC. After performing DFT-based calculations, the atomic displacements between the SCC and NCC for each atom can be expressed as,

$$(\Delta x, \Delta y, \Delta z). \quad\quad Eq.\ 5$$

Correspondingly, the magnitude of the atomic movement is denoted by $\Delta R$.

For Category-2 materials, experimentally measured structures are treated as the SCC, while the NCC is constructed by restoring symmetry. For example, in $YBa_2Cu_3O_7$, atoms in the Cu-O (x-y) plane can be adjusted to the same z level; see details in [8]. The present work focuses on Category-1 materials, which are typically conventional superconductors.

As a special case in the present work, $Nb_3Sn$ ($T_c \approx 18$ K [25]) is unique because both the NCC (cubic) and SCC (tetragonal) phases have been experimentally observed (see Section 3.3). These measurements serve to validate our proposed approach for defining the NCC and SCC.

## 2.2 Details of DFT-based calculations for SCC and NCC

All DFT-based calculations in the present work were performed using the VASP code [31], with ion-electron interactions described by the projector augmented wave (PAW) method [32]. Two X-C functionals were employed: the widely used GGA developed by Perdew-Burke-Ernzerhof (PBE) [33] and the highly accurate meta-GGA functional, $r^2$SCAN [17,34]. In all VASP calculations, electron configurations for each element follow those used by the Materials Project [35] and are listed in Table S1. Supercell sizes were set to 3×3×2 (36 atoms) for hcp elements, 3×3×3 (54 atoms) for bcc elements, and 2×2×2 (64 atoms) for $Nb_3Sn$. The electronic self-consistency criterion was consistently set to $10^{-6}$ eV/atom, based on convergence tests from our previous work [8]. For pure elements, selected $k$-point meshes of 7×7×7 were used, and energy cutoffs ($E_{cut}$) were determined by VASP with using PREC = High (specific values listed in Table S1).

In these calculations, structural relaxations were performed using the Methfessel-Paxton smearing [36] for at least three iterations and the charge density was computed using the tetrahedron method with a Blöchl correction [37]. Isosurface levels ($F_{level}$) for plotting the SCC-NCC charge density differences (SNCDDs) were chosen using VESTA [38] (see Table S1). $F_{level}$

was set near the maximum value that still produces connected SNCDDs; lower values would generate fully connected three-dimensional (3D) tunnels across the material. For Nb$_3$Sn, equilibrium properties of volume ($V_0$), bulk modulus ($B_0$) and its derivative with respect to pressure ($B'$) were fitted using the four-parameter Birch-Murnaghan equation of state (EOS) [39] with inputs derived from the DFT energy-volume data.

Symmetry constraints were fully disabled in all VASP calculations involving pure elements to allow for spontaneous symmetry breaking in the SCCs. For Nb$_3$Sn, the cubic A15 phase is ferromagnetic, while the tetragonal phase is non-magnetic.

### 2.3 Identification of superconductor at 0 K and 0 GPa

A material is classified as a superconductor at 0 K and 0 GPa based on the following criteria [8]:

1) **Metallic nature:** Non-zero electronic DOS at the Fermi level, consistent with BCS theory (see Eq. 1).
2) **Ground-state SCC:** The SCC possesses lower energy than the NCC ($\Delta E_{SN} < 0$). This criterion is omitted in the present work because contemporary X-C functionals cannot reliably capture $\Delta E_{SN}$, which is typically in the meV range and fall within DFT error margins.
3) **Formation of SODTs:** The SCC-NCC charge density difference (SNCDD) forms SODTs, serving as the primary indicator of superconductivity in this work. These SODTs result from regular, systematic atomic displacements between the SCC and NCC.

Thus, the key criteria adopted here are metallicity and the successful formation of SODTs, while the energy-based comparison is acknowledged but excluded due to current functional limitations.

### 3  Results and discussion

Table 1 summarizes the DFT-based results for identifying superconductivity of hcp elements (Be, Mg, Sc, Y, Ti, Zr, Hf, Tc, Re, Ru, Os, and Zn; see Sec. 3.1), bcc elements (V, Nb, Ta, Mo, and W; see Sec. 3.2), and Nb$_3$Sn (Sec. 3.3) at 0 K and 0 GPa, using the criteria defined in Sec. 2.3. Table 1 also details available experimental observations and provides a list of figures illustrating the predicted SNCDDs. Atomic displacements between the NCC and SCC for each material are compiled in Table S2, including the average movements obtained from the DFT calculations along with the mean atomic displacements defined in Eq. 5.

## 3.1 HCP metals

### 3.1.1 Be and Mg

Both alkaline earth metals, beryllium (Be) and magnesium (Mg), crystallize in the hcp structure at ambient pressure and low temperatures [40]. As an example, Figure S 1 (a and b) presents the total charge densities of Be with the NCC and SCC configurations as predicted by r$^2$SCAN. The charge densities are nearly identical for both configurations, showing spherical shapes around each atom. However, Figure S 1 (c and d) illustrates the predicted SNCDDs of Be using r$^2$SCAN, revealing SODTs along the *c*-axis for electrons and the *b*-axis for holes. By analyzing all atomic displacements defined in Eq. 5, we find that only two distinct displacement magnitudes occur, i.e., $\Delta R = 0.0481 \pm 6.6\text{E-}5$ Å and $0.0256 \pm 2.0\text{E-}4$ Å (see Table S2). The two values correspond to the Wyckoff-position sites (2c) of the hcp space group *P6$_3$/mmc*, located at $(\frac{1}{3}, \frac{2}{3}, \frac{1}{4})$ and $(\frac{2}{3}, \frac{1}{3}, \frac{3}{4})$. The extremely small standard deviations of these displacement magnitudes (< 0.0002 Å) indicate that the atomic shifts between the SCC and NCC are highly regular and systematic, ultimately giving rise to the formation of the SODTs.

Notably, the GGA functional predicts that both electron and hole SNCDDs exhibit SODTs along the *b*-axis. Table 1 summarizes the predicted energy difference between the SCC and NCC ($\Delta E_{SN}$, cf. Eq. 3). The values are approximately zero (0.005 meV by GGA) and negative (-2.049 meV by r$^2$SCAN), indicating a slight energetic preference for SCC. Based on the predicted SODTs, we conclude that Be is a superconductor at 0 K and 0 GPa. Experimentally, bulk Be has been observed to exhibit superconductivity at ambient pressure with an extremely low $T_c$ of 26 mK [24,41], as evidenced by changes in magnetic susceptibility [41]. These experimental findings, along with the present DFT predictions, are summarized in Table 1.

Figure 1 shows the predicted SNCDDs of hcp Mg using GGA, which reveal charge density characteristics similar to those of hcp Be. Specifically, Mg exhibits SODTs along the *b*-axis for electrons and the *c*-axis for holes. In contrast, r$^2$SCAN does not predict SODTs for hcp Mg (Figure S 2), instead showing a heavy zigzag SNCDD for electrons and a 3D character for holes.

By analyzing the atomic displacements of hcp Mg (see Eq. 5 and Table S2), the GGA-predicted ΔR values can be grouped into only two distinct displacement magnitudes, exhibiting extremely small standard deviations: 0.0420 ± 6.5E-5 Å and 0.0427 ± 6.1E-4 Å. These very small deviations indicate highly regular and systematic atomic shifts between the SCC and NCC, which in turn facilitate the formation of SODTs. In contrast, the r$^2$SCAN-predicted ΔR values, grouped in the same manner as for the GGA case, display much larger standard deviations: 0.0426 ± 0.0178 Å and 0.0493 ± 0.0184 Å. This broad variation arises from irregular atomic displacements along the z-axis, where Δz = -0.0004 ± 0.0161 Å and -0.0055 ± 0.0156 Å, with standard deviations far exceeding the corresponding mean values (see Table S2). Such irregular atomic displacements produce a strongly zigzag SNCDD, preventing the formation of SODTs.

Table 1 indicates that the predicted $\Delta E_{SN}$ values for Mg are close to zero (0.010 meV by GGA and -0.654 meV by r$^2$SCAN). Based on the GGA-predicted SODTs, we conclude that Mg is a superconductor at 0 K and 0 GPa. Although superconductivity has not been experimentally observed in pure Mg at ambient pressure, evidence from Mg-rich Mg-Cd alloys suggests that pure Mg should be superconducting, with an extrapolated $T_C$ around 0.5 mK [42]. Furthermore, Buhrman and Halperin [43] used the superconducting proximity effect to establish a new low upper bound of 2 mK for the transition temperature of Ag, Au, and Mg. These findings align with the present DFT predictions using GGA, as shown in Table 1.

### 3.1.2 Sc and Y

Both scandium (Sc) and yttrium (Y) are closely associated with the rare earth group and crystallize in the hcp structure at ambient pressure and low temperatures [40]. Figure S 3 illustrates the predicted SNCDDs of Sc using the r$^2$SCAN functional, showing SODTs along the [110] direction. Table S2 shows that all atomic displacements of hcp Sc can be grouped into a single, well-defined displacement magnitude with a small standard deviation ($\Delta R = 0.0444 \pm 0.0017$ Å). This narrow distribution indicates that the atomic shifts between the SCC and NCC are highly regular and systematic, thereby enabling the formation of SODTs. Table 1 shows that the predicted $\Delta E_{SN}$ value is close to zero (0.077 meV by GGA and -0.158 meV by r²SCAN). Based on the r²SCAN-predicted SODTs (with GGA showing similar characteristics), we conclude that Sc is a superconductor at 0 K and 0 GPa. Experimentally, superconductivity has not been observed in Sc at ambient pressure, but it has been reported at high pressures, for example, $T_C$ of 0.34 K at 21 GPa [23,24] and 36 K at 260 GPa [44]. These experimental observations, along with the present DFT results, are summarized in Table 1.

Figure S 4 presents the predicted SNCDDs of Y using r²SCAN, which are similar to those of Sc, showing SODTs along the *b*-axis direction for both electron and hole SNCDDs. Similar to the case of hcp Sc, Table S2 shows that hcp Y exhibits a single, well-defined displacement magnitude with a small standard deviation ($\Delta R = 0.0591 \pm 6.0E-4$ Å), indicating that its atomic shifts are sufficiently regular to support the formation of SODTs. Table 1 indicates that the predicted $\Delta E_{SN}$ values are also close to zero (-0.0097 meV by GGA and 0.0292 meV by r²SCAN). Based on the r²SCAN predictions (with GGA yielding similar results), we conclude that Y is a superconductor at 0 K and 0 GPa. Similar to Sc, superconductivity in Y has not been observed at ambient pressure above 6 mK [45]. However, it has been detected under high pressures, for example, $T_C$ of 0.34 K at 21 GPa [23,24], 17 K at 89 GPa [45], and 20 K at 115 GPa [46]. These experimental findings and the present DFT predictions for Y are summarized in Table 1.

### 3.1.3 Ti, Zr, and Hf

The early transition metals titanium (Ti), zirconium (Zr), and hafnium (Hf) belong to the same group on the periodic table, and crystallize in the hcp structure at ambient pressure and low temperatures [40]. Figure S 5 illustrates the predicted SNCDDs of hcp Ti predicted using r²SCAN, showing SODTs along the *c*-axis direction for both electron and hole SNCDDs. The GGA functional predicts similar characteristics. Table 1 shows that the predicted $\Delta E_{SN}$ is slightly positive (0.012 meV by GGA and 0.039 meV by r²SCAN). Based on these predicted SODTs, we conclude that Ti is a superconductor at 0 K and 0 GPa. Experimentally, Ti has been confirmed as a conventional Type I superconductor at ambient pressure, with a $T_C$ of approximately 0.5 K [23,24]. These observations are consistent with the present DFT predictions, and the results are summarized in Table 1.

The predicted SNCDDs of hcp Zr and hcp Hf using r²SCAN are shown in Figure S 6 and Figure S 7, respectively. Both elements exhibit similar charge density characteristics, displaying SODTs along the *b*-axis direction for both electron and hole SNCDDs. Table 1 shows that the predicted $\Delta E_{SN}$ values are close to zero: slightly negative for Zr (-0.00003 meV) and slightly positive for Hf (0.036 meV) by r²SCAN. Based on these predicted SODTs, we conclude that Zr and Hf are superconductors at 0 K and 0 GPa. Experimentally, superconductivity in Zr and Hf has been confirmed at ambient pressure using magnetic threshold curves measurements [47], with $T_C$ of 0.6 K for Zr and 0.38 K for Hf [23,24]. These findings agree with the present DFT predictions using both GGA and r²SCAN and are summarized in Table 1.

Note that the formation of SODTs in hcp Ti, Zr, and Hf arises from a single, well-defined displacement magnitude with a small standard deviation for each element ($\Delta R = 0.0978 \pm 7.8\text{E-}4$ Å, $0.0524 \pm 1.9\text{E-}4$ Å and $0.0514 \pm 4.0\text{E-}4$ Å, respectively; see Table S2). These narrow distributions demonstrate that the atomic shifts between the SCC and NCC are highly regular and systematic, thereby supporting the formation of SODTs in Ti, Zr, and Hf.

### 3.1.4 Tc, Re, Ru, and Os

Technetium (Tc) is the lightest element without any stable isotopes and crystallizes in an hcp structure. Despite its radioactivity, its superconducting properties have been well-characterized, exhibiting ambient-pressure superconductivity with a relatively high critical temperature ($T_C$ = 8.2 K [23,24]). Unlike Type I superconductors such as Ti and Zr, Tc is a Type II superconductor, allowing partial magnetic field penetration via quantized vortices. Figure S 8 shows the predicted SNCDDs of hcp Tc using r²SCAN, revealing SODTs along the [110] direction formed by connecting ellipsoidal charge densities. Table 1 indicates slightly negative values for the predicted $\Delta E_{SN}$ (-0.020 meV by GGA and -0.028 meV by r²SCAN). Based on the predicted SODTs, we conclude that hcp Tc is a superconductor at 0 K and 0 GPa, consistent with experimental observations [23,24]. Table 1 summarizes the present DFT predictions (using both GGA and r$^2$SCAN) and experimental findings for hcp Tc.

Figure S 9 presents the predicted SNCDDs of hcp rhenium (Re) using both GGA and r$^2$SCAN. Only the hole SNCDD by GGA shows SODT along the *b*-axis direction, while no clear SODTs are observed in other predictions. Table 1 shows slightly negative $\Delta E_{SN}$ values (-0.014 meV by GGA and -0.35 meV by r$^2$SCAN). Based on the GGA-predicted SODTs, we conclude that hcp Re is a superconductor at 0 K and 0 GPa. Experimentally, pure Re is a conventional Type I superconductor at ambient pressure with a measured $T_C$ of 1.7 K [23,24], which agrees with the present DFT predictions. These results are summarized in Table 1.

Figure S 10 illustrates the predicted electron and hole SNCDDs of hcp ruthenium (Ru) by r$^2$SCAN, showing SODTs along the *c*-axis direction. Although the charge density exhibits weak zigzag characteristics, these features do not disrupt the SODTs (similar features are seen in Os and Zn). Table 1 shows that the predicted $\Delta E_{SN}$ values are slightly negative (-0.014 meV by GGA and -0.012 meV by r$^2$SCAN). Based on the r$^2$SCAN-predicted SODTs (with GGA yielding similar results), we conclude that hcp Ru is a superconductor at 0 K and 0 GPa. Experimentally, pure Ru is a conventional Type I superconductor at ambient pressure with a measured $T_C$ of 0.5 K [23,24]. Experimental findings agree with the present DFT predictions, and these results are summarized in Table 1.

Figure S 11 shows the predicted SNCDDs of hcp osmium (Os) using r$^2$SCAN, revealing SODTs along the *b*-axis direction with slight zigzag features for both electron and hole distributions. Table 1 shows slightly positive $\Delta E_{SN}$ values (0.0189 meV by GGA and 0.1611 meV by r$^2$SCAN). Based on the r$^2$SCAN-predicted SODTs (with GGA showing similar results), we conclude that hcp Os is a superconductor at 0 K and 0 GPa, consistent with experimental observations. Pure Os is a conventional Type I superconductor at ambient pressure with a measured $T_C$ of 0.7 K [23,24]. These calculated and measured results are summarized in Table 1.

Consistent with the above analyses, the formation of SODTs in hcp Tc, Ru, and Os obtained using r$^2$SCAN and hcp Re obtained using GGA (instead of r$^2$SCAN), arises from highly regular and systematic atomic shifts between the SCC and NCC (see Table S2). These well-ordered displacements support the formation of SODTs in Tc, Ru, and Os ($r^2$SCAN) and in Re (GGA).

In contrast, for hcp Re calculated using $r^2$SCAN, which mirrors the behavior observed for hcp Mg with $r^2$SCAN, the atomic shifts between the SCC and NCC are irregular, with $\Delta R = 0.0447 \pm 0.0498$ Å and $0.0447 \pm 0.0108$ (see Table S2). The large standard deviations relative to the mean values indicate pronounced structural disorder, which prevents the formation of SODTs in hcp Re under $r^2$SCAN.

### 3.1.5 Zn

Zinc (Zn) crystallizes in the hcp structure at ambient pressure and low temperatures [40]. Figure S 12 illustrates the predicted electron and hole SNCDDs of hcp Zn using $r^2$SCAN, showing the SODTs forming along the *c*-axis for electrons and the *b*-axis for holes due to the highly regular and systematic atomic shifts between the SCC and NCC ($\Delta R = 0.0449 \pm 1.0E-4$ Å and $0.0330 \pm 6.6E-4$ Å, see Table S2). Table 1 presents the predicted $\Delta E_{SN}$ values of hcp Zn, which are close to zero (0.0742 meV by GGA and -1.791 meV by $r^2$SCAN). Based on the $r^2$SCAN-predicted SODTs (with GGA yielding similar results), we conclude that hcp Zn is a superconductor at 0 K and 0 GPa. Experimentally, pure Zn is a conventional Type I superconductor at ambient pressure with a measured $T_C$ of 0.85 K [23,24]. These theoretical predictions and experimental observations are summarized in Table 1.

## 3.2 BCC metals: V, Nb, Ta, Mo, and W

The refractory metals vanadium (V), niobium (Nb), tantalum (Ta), molybdenum (Mo), and tungsten (W) all crystallize in a bcc structure at ambient pressure and low temperatures [40]. At ambient pressure, they are all conventional superconductors, classified as either Type I (e.g., Ta, Mo, and W) or Type II (e.g., V and Nb) [23,24]. Among these five elements, Nb exhibits the highest $T_c$ temperature at ambient pressure (9.25 K), which is the highest among all pure elements [23,24], cf. Table 1. The other elements have lower $T_C$ temperatures: 5.4 K for V, 4.4 K for Ta, 0.92 K for Mo, and 0.01 K for W [23,24] (see Table 1).

Figure 2 presents the r²SCAN-predicted electron and hole SNCDDs of bcc V, revealing clear SODTs along [111] for both electrons and holes. By analyzing the atomic displacements of bcc V (see Eq. 5), we find that all atoms exhibit a single, well-defined displacement vector (Table S2):

$$(\Delta x, \Delta y, \Delta z) = (0.0398 \pm 2.7\text{E-}4 \text{ Å}, -0.0133 \pm 1.8\text{E-}4 \text{ Å}, -0.0067 \pm 1.7\text{E-}4 \text{ Å}), \quad Eq.\ 6$$

with a corresponding magnitude $\Delta R = 0.0425 \pm 3.6\text{E-}4$ Å. The extremely small standard deviations (on the order of 1E-4 Å) indicate that the atomic shifts between the SCC and NCC are highly regular and systematic, thereby enabling the formation of the SODTs. Note that all other bcc elements examined (Nb, Ta, Mo, and W) exhibit similarly regular atomic displacements (Table S2).

In addition, GGA predicts similar charge density features, including SODTs along [111] of bcc V, see Figure S 13. Table 1 shows that the predicted $\Delta E_{SN}$ values are close to zero (0.024 meV by GGA and -1.561 meV by r²SCAN). Based on these predicted SODTs (using both r²SCAN and GGA), we conclude that bcc V is a superconductor at 0 K and 0 GPa, consistent with experimental observations ($T_C$ = 5.4 K [23,24]).

Similar to bcc V, the other refractory metals Nb, Ta, Mo, and W exhibit comparable SNCDDs, as shown in Figure S 14, Figure S 15, Figure S 16, and Figure S 17, respectively, based on r²SCAN predictions (with GGA yielding similar results). All these SNCDDs exhibit SODTs along [111] for both electrons and holes, arising from the highly regular and systematic atomic shifts between the SCC and NCC (Table S2). Table 1 shows that the predicted $\Delta E_{SN}$ values are close to zero ($|\Delta E_{SN}| < 0.06$ meV for both GGA and r²SCAN). Based on these predicted SODTs, we conclude that the bcc Nb, Ta, Mo, and W are superconductor at 0 K and 0 GPa, in agreement with experimental observations [23,24]. All theoretical and experimental results are summarized in Table 1.

### 3.3 Nb₃Sn

Niobium-tin (Nb₃Sn), a member of A15 family of superconductors, crystallizing in a cubic structure with space group $Pm\bar{3}n$ (no. 223) [25,26]. At approximately 43 K, it undergoes a displacive martensitic transformation from the high-temperature cubic phase to a low-temperature tetragonal phase with space group $P4_2/mmc$ (no. 131) [48–50]. This transformation is driven by a Jahn-Teller distortion [49].

Figure 3a shows the relaxed tetragonal configuration at 0 K predicted by GGA, with Sn atoms at Wyckoff site 2c (0, 0.5, 0), Nb1 atoms at 2e (0, 0, 0.25), and Nb2 atoms at 4k (*0.25-δ, 0.5, 0.5*). Experimentally, the lattice parameter ratio $a_0/c_0$ changes from 1 to 1.0062 [50], closely matching the GGA prediction of 1.0087 (see Table 2). The Nb2 atomic displacement $|\delta|$ increases from 0 to 0.0031 at 4 K experimentally [50], compared to 0.0052 predicted by GGA. Below 18 K, tetragonal Nb$_3$Sn becomes superconducting [25]. These observations suggest that superconductivity in Nb$_3$Sn at low temperatures originates from a structural transformation from the high-temperature cubic phase (T > 43 K) to the low-temperature tetragonal phase, where the SCC dominates. This behavior is analogous to BaFe$_2$As$_2$, where a previous zentropy-based study showed that superconductivity emerges in the low-temperature antiferromagnetic (AFM) spin-density-wave (SDW) phase, with a thermal population of the SDW-AFM configuration exceeding 99.99% [51]. Since superconductivity is fragile and can be suppressed by exceeding any critical parameters, such as $T_c$, magnetic field, or current density, a dominant SCC is essential [52,53].

Figure 3b illustrates the energy versus volume (*E-V*) curves of cubic nonmagnetic (NM) and ferromagnetic (FM) configurations, and the tetragonal NM configuration of Nb$_3$Sn, calculated using GGA. The tetragonal FM configuration relaxes to the NM configuration, while the cubic FM phase exhibits an average magnetic moment of 0.106 μ$_B$/atom at equilibrium volume. Figure 3b indicates that the ground state of Nb$_3$Sn is the tetragonal NM configuration with a predicted $\Delta E_{SN}$ = -0.392 meV/atom with respect to the cubic FM configuration (Table 2).

Table 2 summarizes the crystallographic and structural properties of both cubic and tetragonal Nb$_3$Sn predicted by GGA, showing good agreement with experimental data, including equilibrium volume ($V_0$), $a_0/c_0$ ratio, and bulk modulus ($B_0$) [48,54]. Unlike the elemental cases discussed earlier, the high-temperature cubic A15 phase is considered as the NCC, while the low-temperature tetragonal phase is SCC for Nb$_3$Sn. Figure 4 shows the predicted electron and hole SNCDDs of Nb$_3$Sn using GGA. The electron SNCDDs exhibit distinct SODTs along the *c*-axis (Figure 4a), arising from the highly regular and systematic atomic shifts between the experimental SCC and experimental NCC (whose standard derivations are set to zero for these experimental configurations, see Table S2), despite the Nb2 atomic displacements occurring along the *a*- and *b*-axes (Figure 3a).

In contrast, the hole SNCDDs form a 3D network rather than SODTs. Based on the predicted electron SODTs, negative $\Delta E_{SN}$, and metallic nature (absence of bandgap), we conclude that tetragonal $Nb_3Sn$ is a superconductor at 0 K and 0 GPa, due primarily to electron transport along the *c*-axis. These findings are summarized in Table 1, align with experimental observations that tetragonal $Nb_3Sn$ becomes superconducting below $T_C \approx 18$ K [25]. In the tetragonal phase, superconducting behavior is complex and no preferred crystallographic direction has been experimentally identified [55].

## 4 Summary

The present work verifies a recently proposed unified theoretical framework for superconductivity applicable to both conventional and unconventional materials [8]. The theoretical framework suggests that superconductivity originates from the formation of a *symmetry-broken* superconducting configuration (SCC), created by atomic perturbations of the normal conducting configuration (NCC). This transformation, driven by electron-phonon interactions, results in straight one-dimensional tunnels (SODTs) in the charge density difference, arising from the highly regular and systematic atomic shifts between the SCC and NCC. The verification is through DFT-based calculations at 0 K and 0 GPa for 12 hcp elements (Be, Mg, Sc, Y, Ti, Zr, Hf, Tc, Re, Ru, Os, and Zn), 5 bcc elements (V, Nb, Ta, Mo, and W); and one compound $Nb_3Sn$. Our results indicate that all these materials exhibit SODTs, supporting their superconducting nature at 0 K and 0 GPa. However, the superconductivity of Mg, Sc, and Y remain to be confirmed by experiments.

Current DFT calculations face limitations in accurately resolving the energy difference ($\Delta E_{SN}$) between the SCC and NCC due to approximations in X-C functionals, highlighting the need for further methodological improvements. Once the SCC is identified for a given material, its superconducting transition temperature ($T_C$) may be estimated by analyzing the competition between the SCC and NCC. This can be achieved using frameworks such as zentropy theory [21,22,56], which employs coarse-graining of entropy spanning multiple scales from macroscopic observations down to quantum-level fluctuations.

## 5 Acknowledgements

This work was supported by the U.S. Department of Energy (DOE) through Grant No. DE-SC0023185. First-principles calculations were performed partially on the Roar supercomputers at the PSU's Institute for Computational and Data Sciences (ICDS), partially on the resources of the National Energy Research Scientific Computing Center (NERSC), a DOE Office of Science User Facility supported under Contract No. DE-AC02-05CH11231 using the NERSC award BES-ERCAP0032760, and partially on the resources of Advanced Cyberinfrastructure Coordination Ecosystem: Services & Support (ACCESS) through allocation DMR1400063, which is supported by U.S. National Science Foundation Grants Nos. 2138259, 2138286, 2138307, 2137603, and 2138296.

## 6 Data availability statement

All data and plots that support the findings of the present work are included within this article and its supplementary files.

## 7 ORCID IDs

Zi-Kui Liu  https://orcid.org/0000-0003-3346-3696
Shun-Li Shang  https://orcid.org/0000-0002-6524-8897

# 8 Tables and Table Captions

Table 1. DFT-based predictions for superconductivity and related properties of pure elements and compounds: Superconductivity (Y for Yes, N for No), electrical conductivity (Cond: Y for conductor, N for nonconductor), whether SCC is the ground state (GS: Y or N), the energy difference between SCC and NCC ($\Delta E_{SN}$ in meV/atom), the morphology of the SCC-NCC charge density difference (SNCDD), and the predicted versus experimental superconductivity ($Y_0$ at zero pressure and $Y_h$ at high pressure).

| Mater. | X-C | DFT results | | | Characteristics of SNCDD by DFT | Superconductivity | | Figures |
|---|---|---|---|---|---|---|---|---|
| | | Cond | GS | $\Delta E_{SN}$ | | Calc.[a] | Expt.[b] | |
| Be | GGA | Y | N | 0.005 | SODT along $c$- or $b$-axis. | $Y_0$ | $Y_0$ (Type I) 26 mK | Figure S 1 |
| | r$^2$SCAN | Y | Y | -2.049 | | | | |
| Mg | GGA | Y | N | 0.010 | SODT along $c$- or $b$-axis by GGA (zigzag electron by r$^2$SCAN, hole 3D) | $Y_0$ | 0.5 mK [c] 2 mK [d] | Figure 1 Figure S 2 |
| | r$^2$SCAN | Y | Y | -0.654 | | | | |
| Sc | GGA | Y | N | 0.077 | SODT along [110] | $Y_0$ | $Y_h$ 0.34 K at 21 GPa | Figure S 3 |
| | r$^2$SCAN | Y | Y | -0.158 | | | | |
| Y | GGA | Y | Y | -0.0097 | SODT along $b$-axis. | $Y_0$ | $Y_h$ 2.8 K at 15 GPa | Figure S 4 |
| | r$^2$SCAN | Y | N | 0.0292 | | | | |
| Ti | GGA | Y | N | 0.012 | SODT along $c$-axis. | $Y_0$ | $Y_0$ (Type I) 0.5 K | Figure S 5 |
| | r$^2$SCAN | Y | N | 0.039 | | | | |
| Zr | GGA | Y | Y | -0.0006 | SODT along $b$-axis. | $Y_0$ | $Y_0$ (Type I) 0.6 K | Figure S 6 |
| | r$^2$SCAN | Y | Y | -0.00003 | | | | |
| Hf | GGA | Y | N | 0.010 | SODT along $b$-axis. | $Y_0$ | $Y_0$ (Type I) 0.38 K | Figure S 7 |
| | r$^2$SCAN | Y | N | 0.036 | | | | |
| Tc | GGA | Y | Y | -0.020 | SODT along [110] | $Y_0$ | $Y_0$ (Type II) 8.2 K | Figure S 8 |
| | r$^2$SCAN | Y | Y | -0.028 | | | | |
| Re | GGA | Y | Y | -0.014 | SODT along $b$-axis by GGA | $Y_0$ | $Y_0$ (Type I) 1.7 K | Figure S 9 |
| | r$^2$SCAN | Y | Y | -0.35 | | | | |
| Ru | GGA | Y | Y | -0.014 | SODT along $c$-axis (with zigzag character). | $Y_0$ | $Y_0$ (Type I) 0.5 K | Figure S 10 |
| | r$^2$SCAN | Y | Y | -0.012 | | | | |
| Os | GGA | Y | N | 0.0189 | SODT along $b$-axis. | $Y_0$ | $Y_0$ (Type I) 0.7 K | Figure S 11 |
| | r$^2$SCAN | Y | N | 0.1611 | | | | |
| Zn | GGA | Y | N | 0.0742 | SODT along $c$- or $b$-axis. | $Y_0$ | $Y_0$ (Type I) 0.85 K | Figure S 12 |
| | r$^2$SCAN | Y | Y | -1.791 | | | | |
| V | GGA | Y | N | 0.024 | SODT along [111] | $Y_0$ | $Y_0$ (Type II) 5.4 K | Figure S 13 Figure 2 |
| | r$^2$SCAN | Y | Y | -1.561 | | | | |
| Nb | GGA | Y | Y | -0.009 | SODT along [$\bar{1}$11] | $Y_0$ | $Y_0$ (Type II) 9.25 K | Figure S 14 |
| | r$^2$SCAN | Y | N | 0.056 | | | | |
| Ta | GGA | Y | Y | -0.009 | SODT along [111] | $Y_0$ | $Y_0$ (Type I) 4.4 K | Figure S 15 |
| | r$^2$SCAN | Y | Y | -0.039 | | | | |
| Mo | GGA | Y | N | 0.003 | SODT along [$\bar{1}\bar{1}$1] | $Y_0$ | $Y_0$ (Type I) 0.92 K | Figure S 16 |
| | r$^2$SCAN | Y | N | 0.030 | | | | |
| W | GGA | Y | Y | -0.009 | SODT along [111] | $Y_0$ | $Y_0$ (Type I) 0.01 K | Figure S 17 |
| | r$^2$SCAN | Y | Y | -0.002 | | | | |
| Nb$_3$Sn | GGA | Y | Y | -0.392 | SODT along $c$-axis for electrons. | $Y_0$ | $Y_0$ (Type II) 17.9 K | Figure 4 |
| | r$^2$SCAN | Y | Y | -1.003 | | | | |

[a] This work with $Y_0$ for superconductor at zero pressure.

[b] Experimentally observed superconducting elements at ambient pressure (marked by $Y_0$) or high pressure (marked by $Y_h$), and the values indicate the measured $T_c$ for pure elements [23,24] and Nb$_3$Sn [25].

[c] Extrapolation from the data on Mg-rich alloys gives a value of Tc = 0.5 mK for pure magnesium [42].

[d] Using the proximity effect, the established a new upper bound of 2 mK for the transition temperature of Ag, Au, and most importantly Mg [43].

Table 2. Crystallographic and structural properties of cubic and tetragonal Nb$_3$Sn by DFT-based calculations using GGA in comparison with experiments.

| Note | $E_0$ (meV/atom) | $V_0$ (Å$^3$/atom) | $a_0/c_0$ | $B_0$ (GPa) | $B'$ | MM ($\mu_B$/atom) |
|---|---|---|---|---|---|---|
| DFT, cubic (FM) | 0.000 | 18.961 | 1 | 163.5 | 4.27 | 0.106 |
| DFT, cubic (NM) | 0.262 | 18.958 | 1 | 163.6 | 4.29 | 0 |
| DFT, tet (NM) | -0.392 | 18.964 | 1.0087 | 163.3 | 4.33 | 0 |
| Expt., cubic |  | 18.42 [a] | 1 | 165 [b] |  |  |
| Expt., tet |  | 18.43 [a] | 1.0061 [a] | 161 [b] |  |  |

[a] Measured at 5.1 K for tetragonal phase and 50 K for cubic phase [48].

[b] Used by Lu and Klein [54].

## 9 Figures and Captions (main)

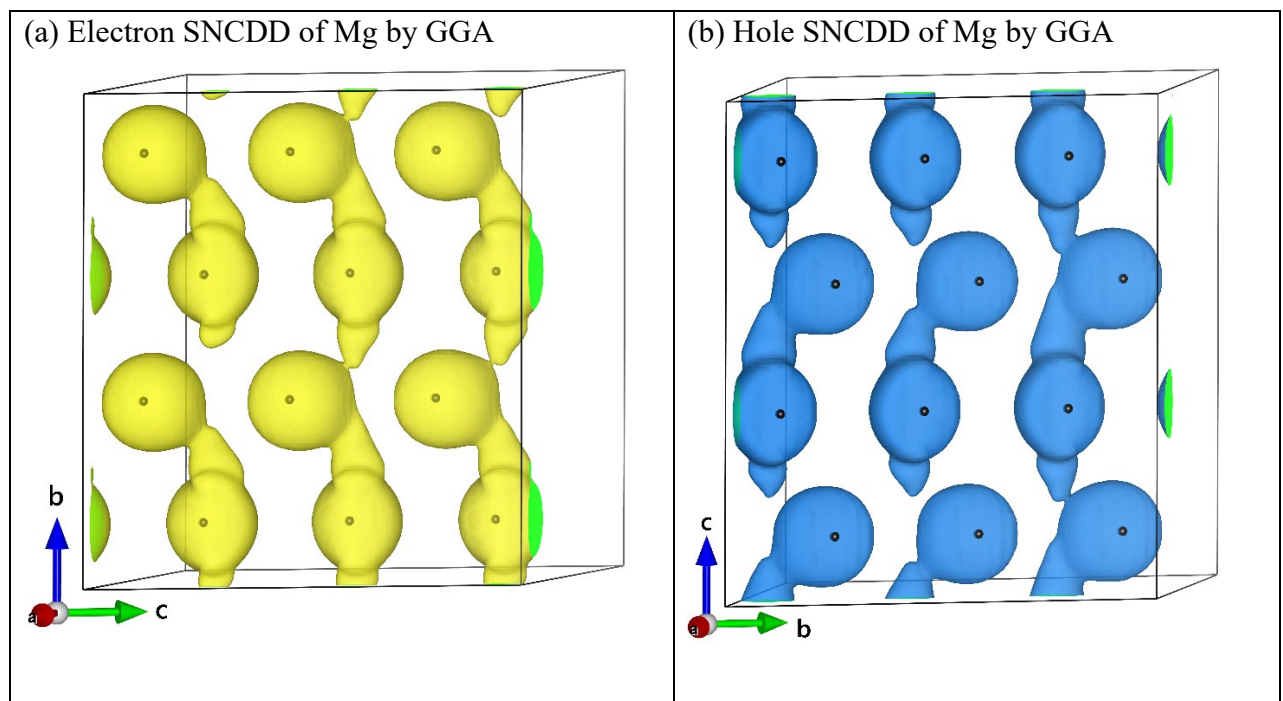

*Figure 1. Partial electron (a, yellow) and hole (b, blue) SNCDDs of Mg predicted by GGA-PBE, showing SODTs along the b-axis for electron and along the c-axis for hole SNCDD. In contrast, r²SCAN does not exhibit SODTs for Mg, see Figure S 2.*

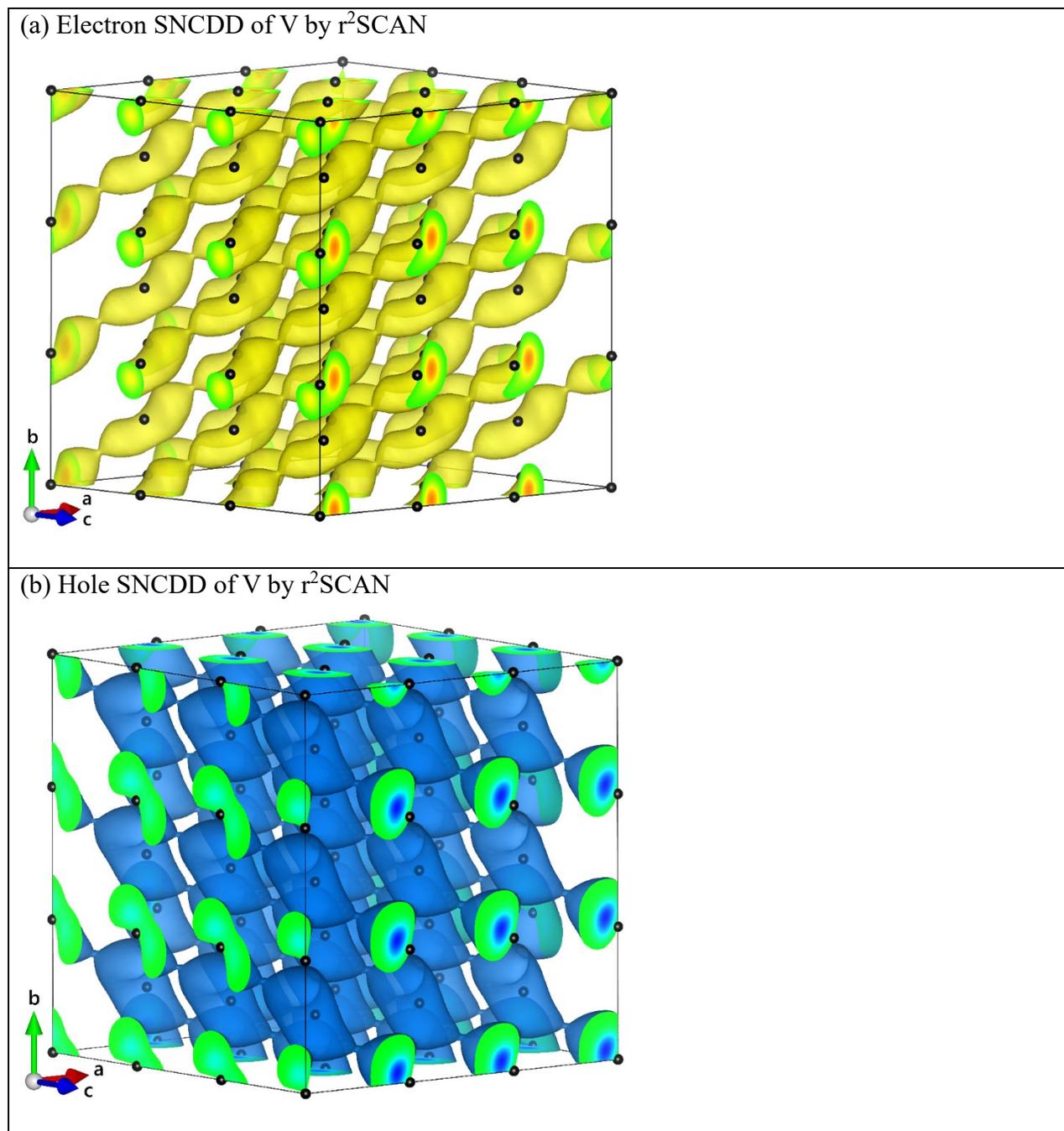

*Figure 2. Electron (a, yellow) and hole (b, blue) SNCDDs of V predicted by r²SCAN, showing SODTs along [111]. For comparison, predictions by GGA-PBE are provided in Figure S 13.*

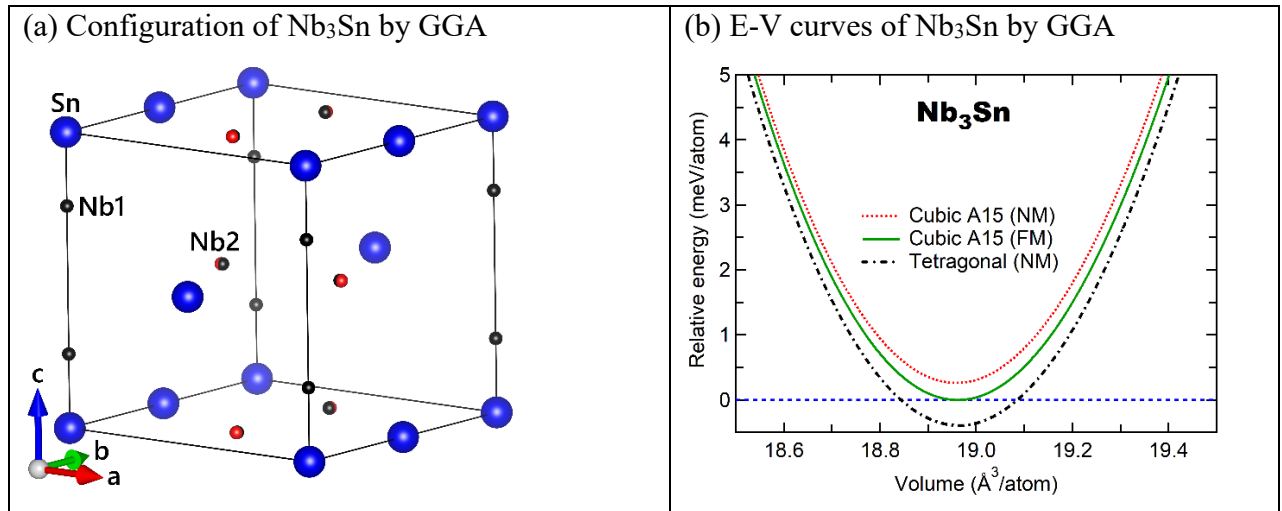

*Figure 3. Relaxed crystal configuration (a) and the predicted energy-volume (E-V) curves for Nib$_3$Sn using GGA-PBE. For the tetragonal Nb$_3$Sn, the Nb2 atoms (red plus black) at Wyckoff site 4m (0.25-δ, 0.25, 0) exhibit a displacement |δ| around 0.0043 according to GGA predictions, compared to the measured |δ| = 0.0031 at 4 K [50].*

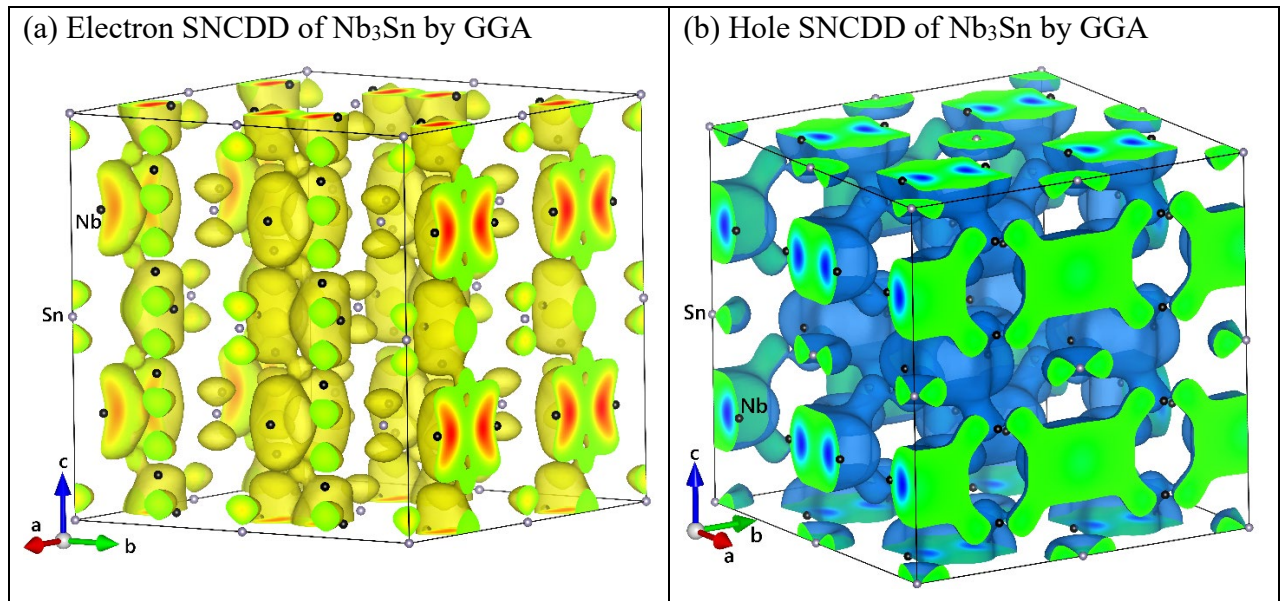

*Figure 4. Electron (a, yellow) and hole (b, blue) SNCDDs of Nb$_3$Sn predicted by GGA-PBE, showing SODTs along the c-axis for electron, while no SODTs for hole.*

# Supplementary Material

# Symmetry-broken superconducting configurations from density functional theory for bcc and hcp metals and Nb$_3$Sn

Shun-Li Shang and Zi-Kui Liu

Department of Materials Science and Engineering, The Pennsylvania State University, University Park, PA 16802, USA

## 1 Supplementary Tables

Table S 1. Settings used to generate SCCs, perform DFT-based calculations, and plot electron SNCDDs, including the crystal structures (str.), the initial atomic perturbation $\delta_{ini}$ (cf., Eq. 4), X-C functionals, $k$-point meshes, cutoff energy ($E_{cut}$ in eV, determined by the VASP setting "PREC = High"), the minimum ($F_{min}$) and maximum ($F_{max}$) charge density differences, and the contour levels for plotting electron SNCDDs ($F_{level}$) with charge gain.

| Materials | Str. | $\delta_{ini}$ | X-C | $k$-mesh | $E_{cut}$ | $F_{min}$ | $F_{max}$ | $F_{level}$ | Figure |
|---|---|---|---|---|---|---|---|---|---|
| Be_sv (4) [a] | hcp | 0.04 | r$^2$SCAN | 7×7×7 | 401.4 | -0.300982 | 0.316675 | 0.000314 | Figure S 1 |
| Mg_pv (8) [a] | hcp | 0.06 | PBE | 7×7×7 | 525.1 | -0.213652 | 0.214463 | 0.0001408 | Figure 1 |
|  |  |  | r$^2$SCAN |  |  | -0.320174 | 0.313314 | 0.00021 | Figure S 2 |
| Sc_sv (11) [a] | hcp | 0.02 | r$^2$SCAN | 7×7×7 | 289.5 | -0.112994 | 0.111195 | 0.00018 | Figure S 3 |
| Y_sv (11) [a] | hcp | 0.04 | r$^2$SCAN | 7×7×7 | 263.4 | -0.0605017 | 0.0599861 | 0.000188 | Figure S 4 |
| Ti_pv (10) [a] | hcp | 0.08 | r$^2$SCAN | 7×7×7 | 289.0 | -0.204697 | 0.204974 | 0.00061 | Figure S 5 |
| Zr_sv (12) [a] | hcp | 0.04 | r$^2$SCAN | 7×7×7 | 298.9 | -0.0701622 | 0.069794 | 0.00048 | Figure S 6 |
| Hf_pv (10) [a] | hcp | 0.04 | r$^2$SCAN | 7×7×7 | 286.4 | -0.0450728 | 0.0442788 | 0.00058 | Figure S 7 |
| Tc_pv (13) [a] | hcp | 0.02 | r$^2$SCAN | 7×7×7 | 342.6 | -0.0663961 | 0.0665165 | 0.0011 | Figure S 8 |
| Re_pv (13) [a] | hcp | 0.04 | r$^2$SCAN | 7×7×7 | 294.1 | -0.110753 | 0.109115 | 0.00325 | Figure S 9 |
| Ru_pv (14) [a] | hcp | 0.06 | r$^2$SCAN | 7×7×7 | 312.1 | -0.0729848 | 0.0731499 | 0.00138 | Figure S 10 |
| Os_pv (14) [a] | hcp | 0.04 | r$^2$SCAN | 7×7×7 | 296.4 | -0.0748346 | 0.0747532 | 0.00182 | Figure S 11 |
| Zn (12) [a] | hcp | 0.06 | r$^2$SCAN | 7×7×7 | 359.7 | -0.248358 | 0.241290 | 0.00078 | Figure S 12 |
| V_pv (11) [a] | bcc | 0.04 | PBE | 7×7×7 | 342.8 | -0.143164 | 0.138092 | 0.00031 | Figure S 13 |
|  |  | 0.04 | r$^2$SCAN |  |  | -0.179615 | 0.186421 | 0.014 | Figure 2 |
| Nb_pv (11) [a] | bcc | 0.02 | r$^2$SCAN | 7×7×7 | 271.2 | -0.0171389 | 0.0169606 | 0.00019 | Figure S 14 |
| Ta_pv (11) [a] | bcc | 0.04 | r$^2$SCAN | 7×7×7 | 290.8 | -0.0445001 | 0.0445570 | 0.000425 | Figure S 15 |
| Mo_pv (12) [a] | bcc | 0.02 | r$^2$SCAN | 7×7×7 | 292.0 | -0.0217756 | 0.0218054 | 0.00042 | Figure S 16 |
| W_sv (14) [a] | bcc | 0.04 | r$^2$SCAN | 7×7×7 | 290.0 | -0.0837359 | 0.0838688 | 0.00095 | Figure S 17 |
| Nb$_3$Sn [b] | A15 |  | PBE | 7×7×7 | 313.4 | -0.0233904 | 0.0181413 | 0.0021 | Figure 4 |

[a] Valence electrons used in the present DFT calculations. In addition, the suffixes "_sv, _pv, or _d" after the symbols of some elements indicate the s, p, and d states are considered as valence states.
[b] For element Sn_d, 14 valence electrons were included.

Table S 2. Atomic displacements between the NCC and SCC, including the crystal structures (str.), the X-C functionals, the average atomic displacement vector ($\Delta x$, $\Delta y$, $\Delta z$), and the corresponding magnitude of atomic movement $\Delta R$ (see Eq. 5). Values following the $\pm$ symbol donate the standard deviation $\sigma$.

| Materials | Str. | X-C | $\Delta R \pm \sigma$ (Å) | $\Delta x \pm \sigma$ (Å) | $\Delta y \pm \sigma$ (Å) | $\Delta z \pm \sigma$ (Å) |
|---|---|---|---|---|---|---|
| Be | hcp | r$^2$SCAN | 0.0481 ± 6.6E-5 | 0.0398 ± 5.8E-5 | -0.0204 ± 1.1E-5 | 0.0176 ± 3.0E-5 |
|  |  |  | 0.0256 ± 2.0E-4 | 0.0205 ± 1.2E-4 | -0.0057 ± 1.5E-4 | 0.0142 ± 0.2E-4 |
| Mg | hcp | PBE | 0.0420 ± 6.5E-5 | 0.0400 ± 3.9E-5 | -0.0128 ± 0.6E-5 | 0.0004 ± 5.2E-5 |
|  |  |  | 0.0427 ± 6.1E-4 | 0.0400 ± 1.2E-5 | -0.0149 ± 0.3E-5 | -0.0004 ± 5.1E-5 |
|  |  | r$^2$SCAN | *0.0426 ± 0.0178* | *0.0399 ± 0.0057* | *-0.0149 ± 0.0047* | *-0.0004 ± 0.0161* |
|  |  |  | *0.0493 ± 0.0184* | *0.0478 ± 0.0095* | *-0.0108 ± 0.0018* | *-0.0055 ± 0.0156* |
| Sc | hcp | r$^2$SCAN | 0.0444 ± 17E-4 | 0.0028 ± 0.0013 | -0.0336 ± 0.0006 | -0.0288 ± 0.0010 |
| Y | hcp | r$^2$SCAN | 0.0591 ± 6.0E-4 | 0.0490 ± 4.0E-4 | -0.0212 ± 1.2E-4 | 0.0252 ± 4.4E-4 |
| Ti | hcp | r$^2$SCAN | 0.0978 ± 7.8E-4 | 0.0489 ± 4.4E-4 | -0.0847 ± 0.3E-5 | 0.0000 ± 6.5E-4 |
| Zr | hcp | r$^2$SCAN | 0.0524 ± 1.9E-4 | 0.0432 ± 1.4E-4 | -0.0187 ± 0.9E-4 | 0.0229 ± 0.9E-4 |
| Hf | hcp | r$^2$SCAN | 0.0514 ± 4.0E-4 | 0.0425 ± 3.0E-4 | -0.0184 ± 1.9E-4 | -0.0223 ± 1.9E-4 |
| Tc | hcp | r$^2$SCAN | 0.0369 ± 0.0012 | 0.0023 ± 0.0003 | -0.0276 ± 0.0000 | -0.0243 ± 0.0011 |
| Re | hcp | PBE | 0.0450 ± 9.6E-4 | 0.0370 ± 2.6E-4 | -0.0160 ± 1.0E-4 | 0.0199 ± 9.2E-4 |
|  | hcp | r$^2$SCAN | *0.0447 ± 0.0498* | *0.0368 ± 0.0001* | *-0.0160 ± 0.0000* | *0.0198 ± 0.0498* |
|  |  |  | *0.0447 ± 0.0108* | *0.0368 ± 0.0108* | *-0.0158 ± 0.0000* | *0.0199 ± 0.0005* |
| Ru | hcp | r$^2$SCAN | 0.0358 ± 6.5E-5 | 0.0338 ± 5.4E-5 | -0.0117 ± 0.1E-5 | 0.0000 ± 3.7E-5 |
| Os | hcp | r$^2$SCAN | 0.0442 ± 6.0E-5 | 0.0366 ± 4.2E-5 | -0.0158 ± 1.0E-5 | 0.0192 ± 4.1E-5 |
| Zn | hcp | r$^2$SCAN | 0.0449 ± 1.0E-4 | 0.0331 ± 8.5E-5 | -0.0303 ± 0.4E-5 | 0.0000 ± 5.3E-5 |
|  |  |  | 0.0330 ± 6.6E-4 | 0.0321 ± 0.7E-4 | 0.0077 ± 0.1E-4 | 0.0000 ± 6.5E-4 |
| V | bcc | PBE | 0.0426 ± 11E-4 | 0.0400 ± 7.5E-4 | -0.0133 ± 8.5E-4 | -0.0067 ± 0.7E-4 |
|  |  | r$^2$SCAN | 0.0425 ± 3.6E-4 | 0.0398 ± 2.7E-4 | -0.0133 ± 1.8E-4 | -0.0067 ± 1.7E-4 |
| Nb | bcc | r$^2$SCAN | 0.0180 ± 1.8E-4 | -0.0147 ± 1.0E-4 | 0.0074 ± 1.0E-4 | -0.0074 ± 1.1E-4 |
| Ta | bcc | r$^2$SCAN | 0.0470 ± 2.1E-4 | 0.0441 ± 0.4E-4 | -0.0147 ± 1.5E-4 | -0.0073 ± 1.4E-4 |
| Mo | bcc | r$^2$SCAN | 0.0172 ± 6.4E-5 | -0.0140 ± 2.1E-5 | 0.0070 ± 4.0E-5 | -0.0070 ± 4.5E-5 |
| W | bcc | r$^2$SCAN | 0.0451 ± 2.6E-4 | 0.0423 ± 1.5E-4 | -0.0141 ± 1.5E-4 | -0.0070 ± 1.5E-4 |
| Nb$_3$Sn [a] | A15 | PBE | 0.0233 | 0.0233 | 0 | 0 |

[a] Displacement is only for Wyckoff site 4j (x, 0, 0).

## 2 Supplementary Figures

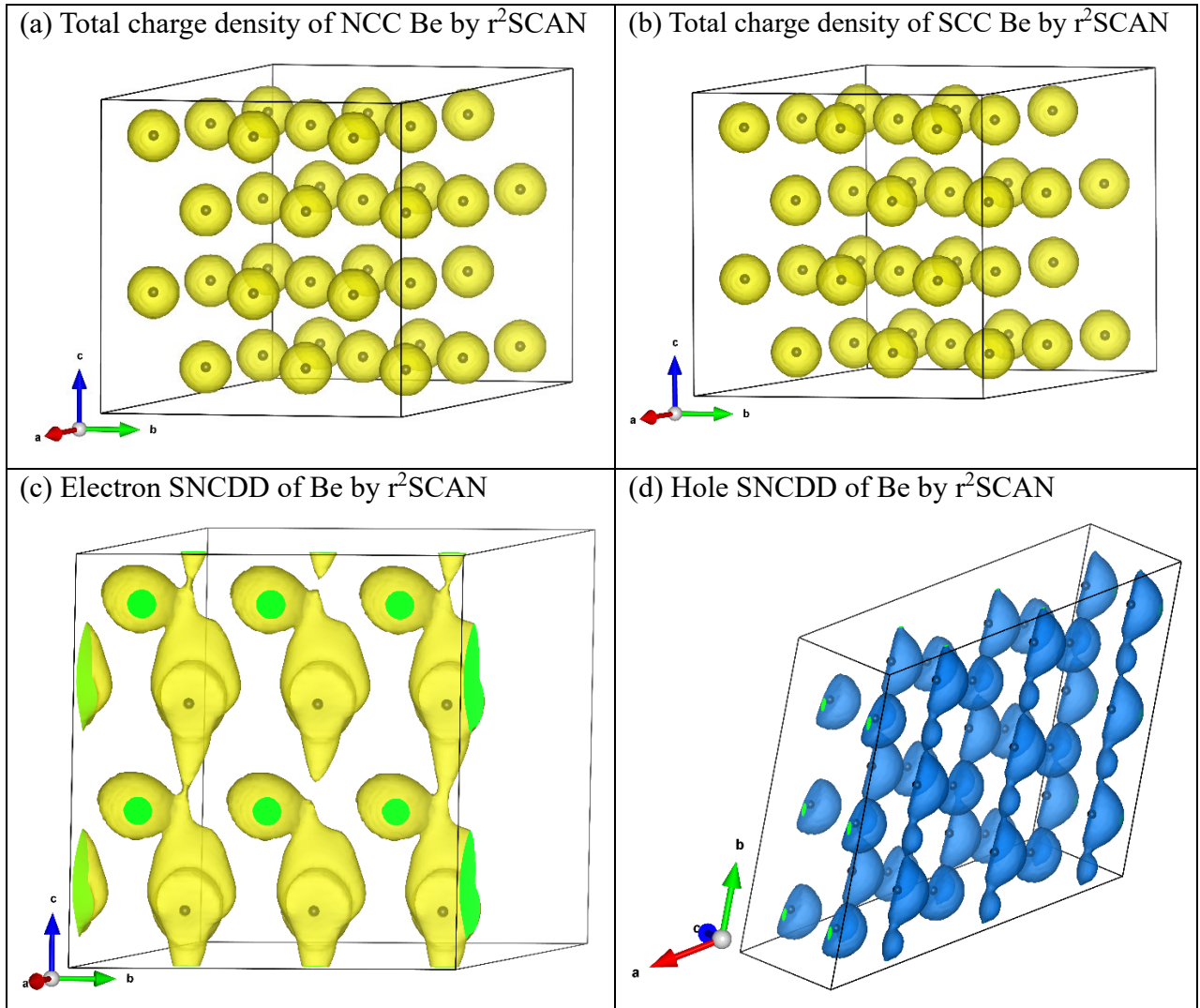

*Figure S 1. (a) Total charge density of Be with NCC configuration and (b) total charge density of Be with SCC configuration. Partial electron (c, yellow) and hole (d, blue) SNCDDs of Be predicted by r²SCAN, showing the SODTs along the c-axis for electron (c) and the b-axis for hole (d). In contrast, GGA-PBE predicts that both electron and hole SNCDDs show SODTs along the b-axis direction (figures not shown).*

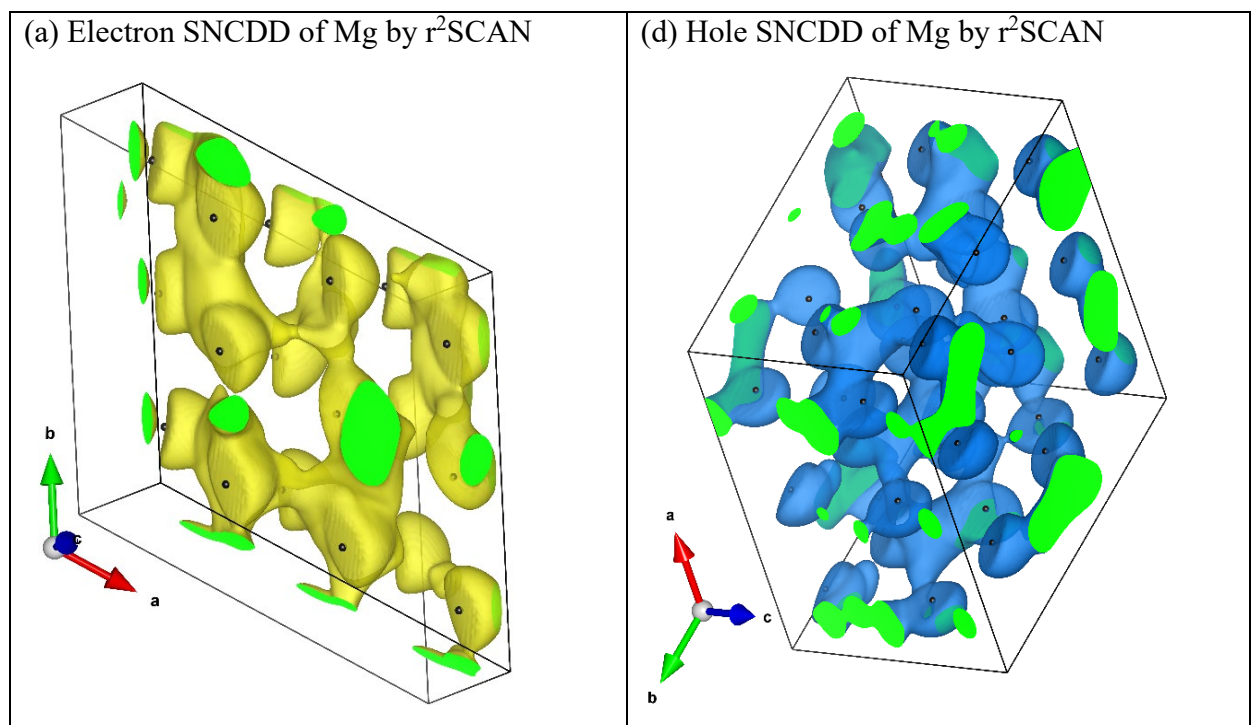

*Figure S 2. Partial electron (a, yellow) and hole (b, blue) SNCDDs of Mg predicted by GGA-r²SCAN, showing a pronounced zigzag pattern for the electron SNCDD for electron and a 3D characteristics for the hole SNCDD; see the predictions by GGA-PBE in Figure 1.*

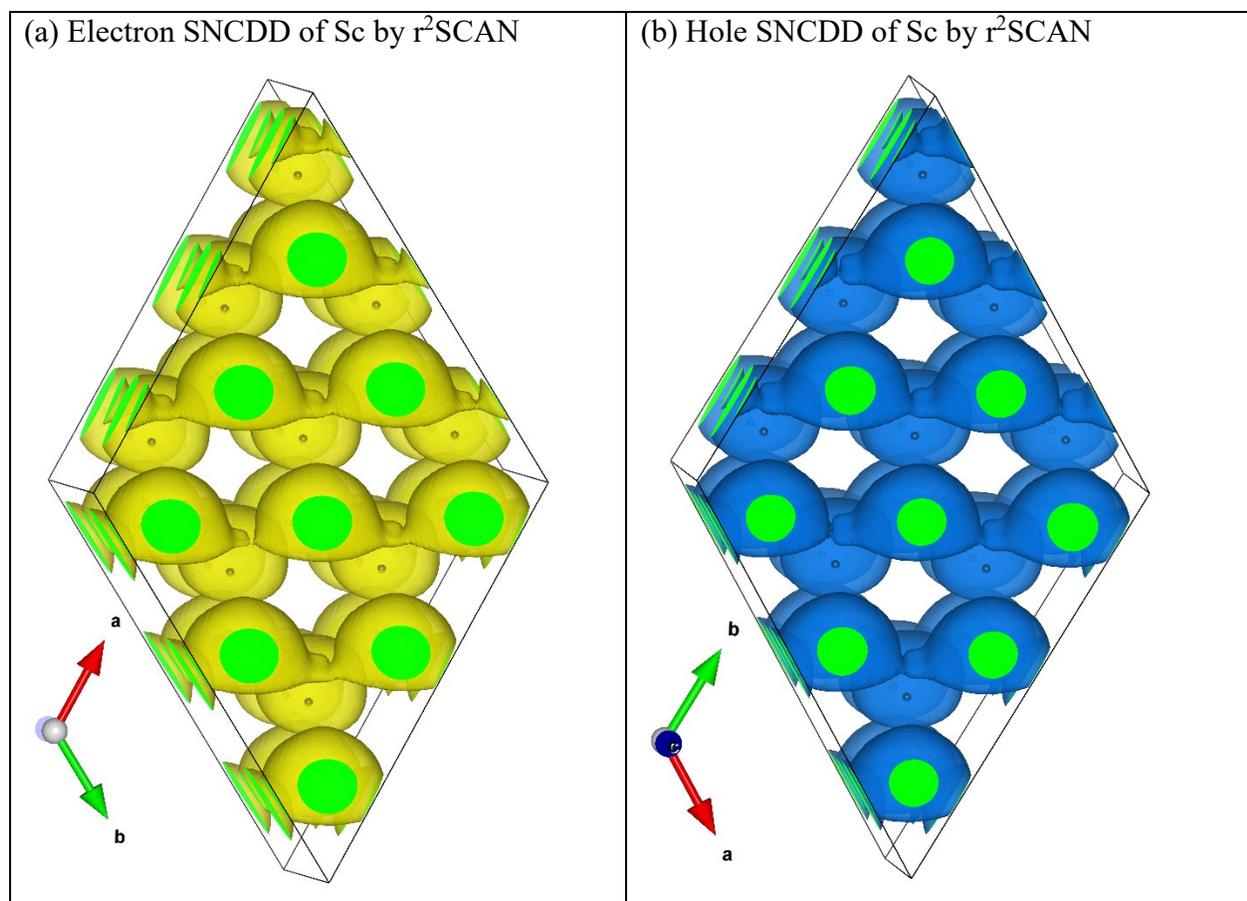

*Figure S 3. Electron (a, yellow) and hole (b, blue) SNCDDs of Sc predicted by r$^2$SCAN, showing SODTs along [110]. Note that GGA-PBE predicts similar characteristics (not shown).*

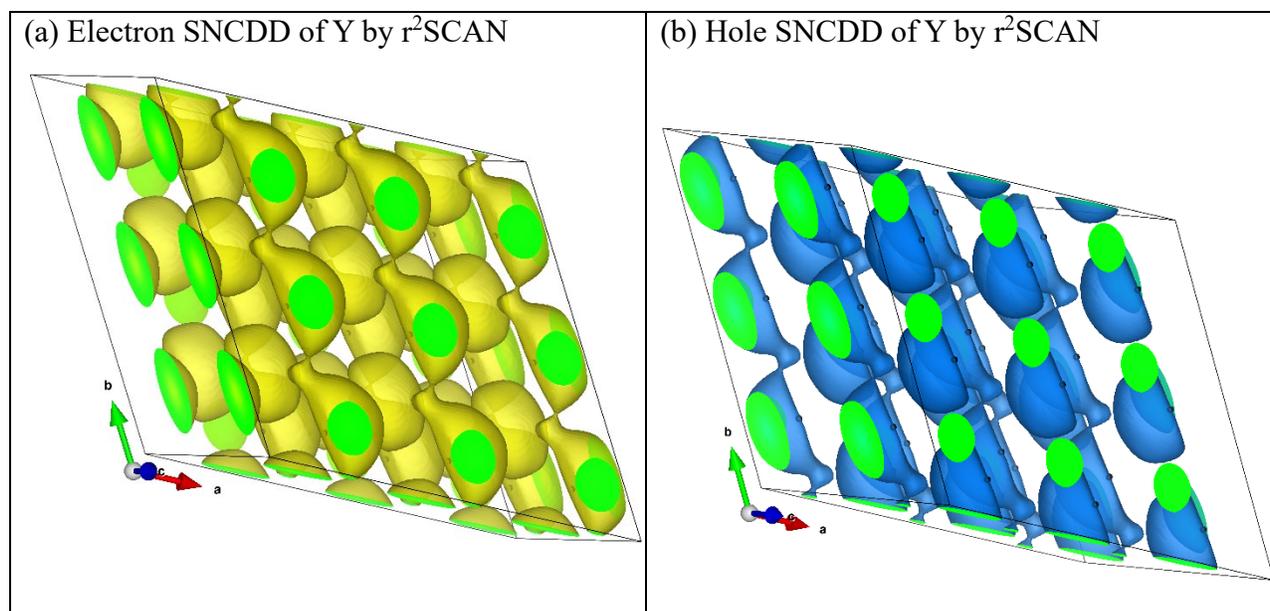

*Figure S 4. Electron (a, yellow) and hole (b, blue) SNCDDs of Y predicted by r$^2$SCAN, showing SODTs along the b-axis. Note that GGA-PBE predicts similar characteristics (not shown).*

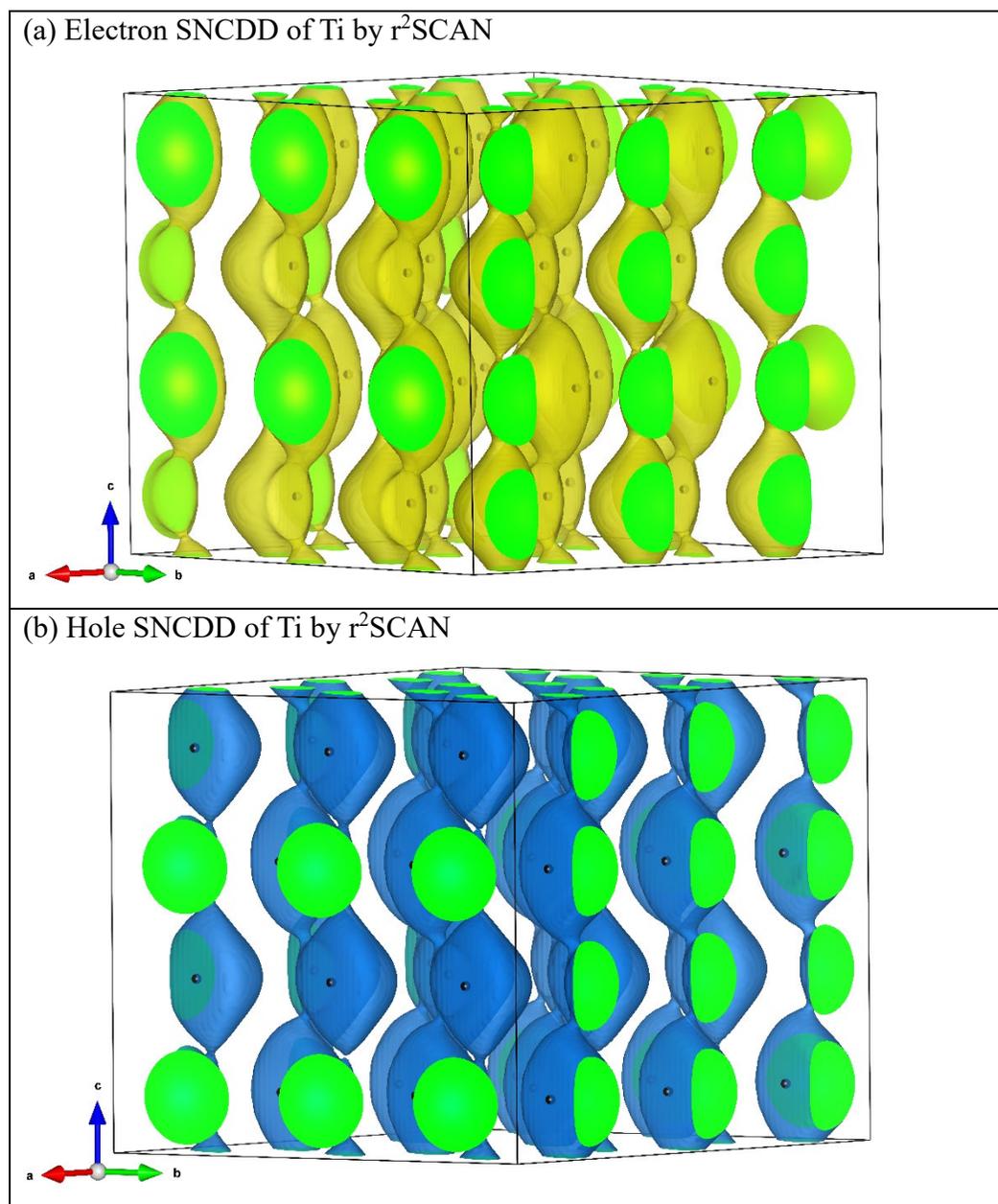

*Figure S 5. Electron (a, yellow) and hole (b, blue) SNCDDs of Ti predicted by $r^2$SCAN, showing SODTs along the c-axis. Note that GGA-PBE predicts similar characteristics (not shown).*

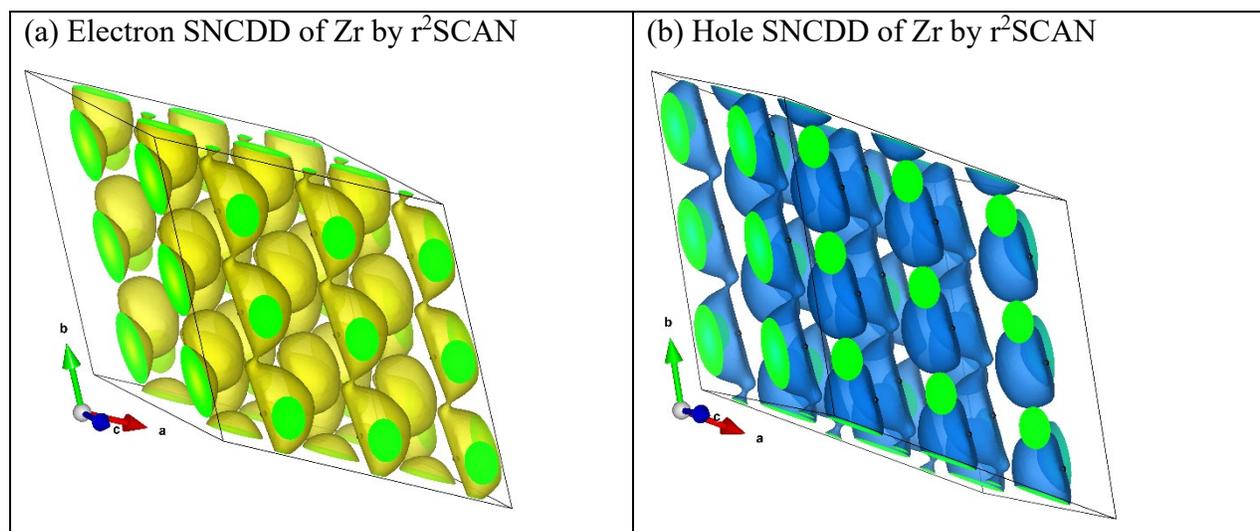

*Figure S 6. Electron (a, yellow) and hole (b, blue) SNCDDs of Zr predicted by r$^2$SCAN, showing SODTs along the b-axis. Note that GGA-PBE predicts similar characteristics (not shown).*

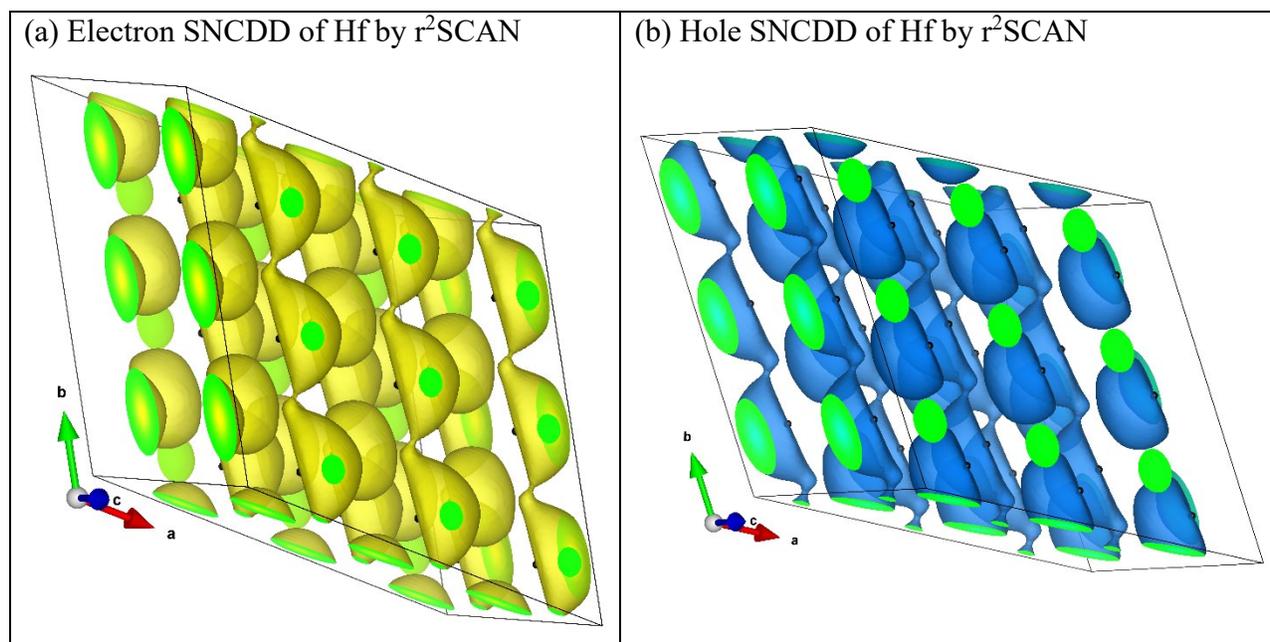

*Figure S 7. Electron (a, yellow) and hole (b, blue) SNCDDs of Hf predicted by $r^2$SCAN, showing SODTs along the b-axis. Note that GGA-PBE predicts similar characteristics (not shown).*

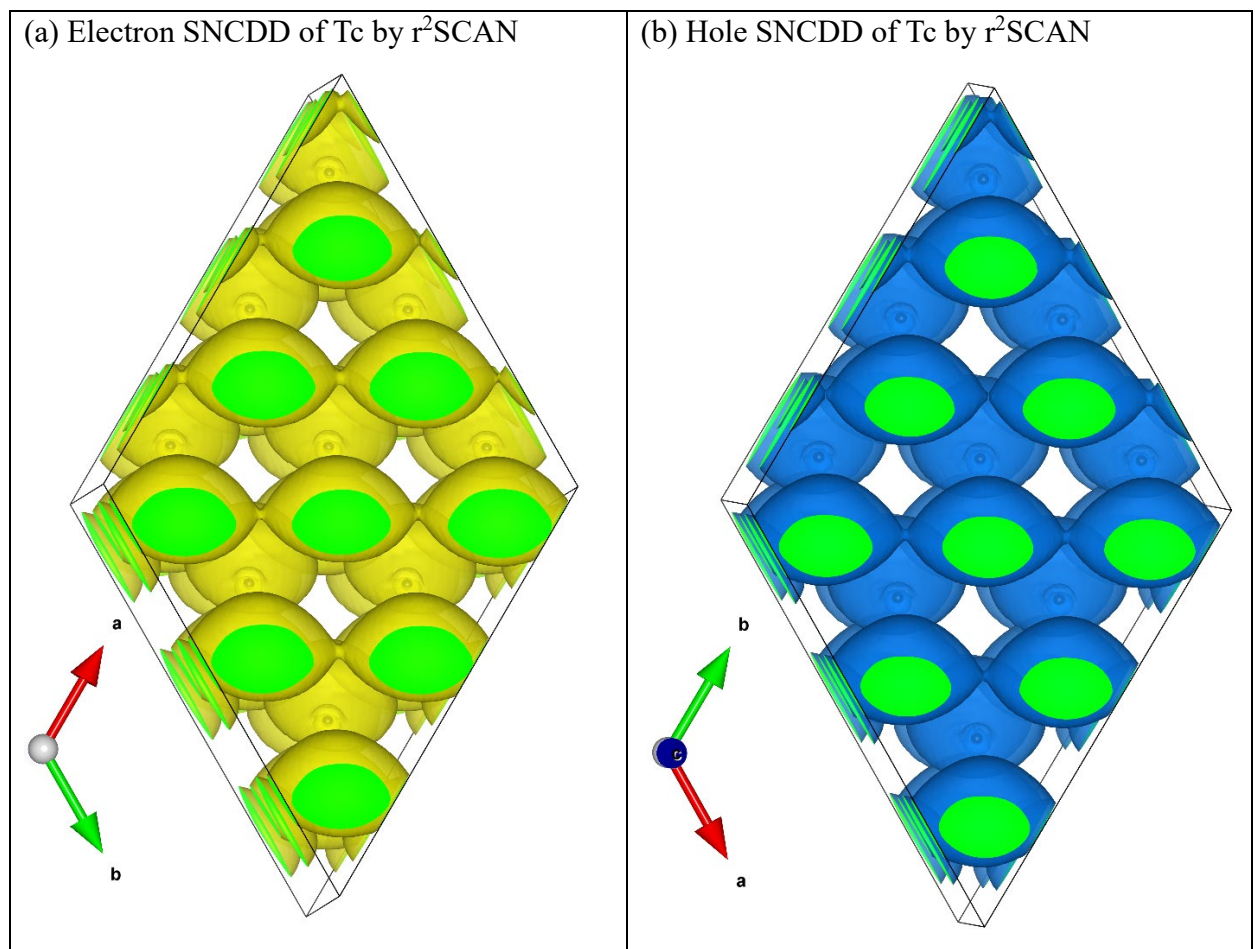

*Figure S 8. Electron (a, yellow) and hole (b, blue) SNCDDs of Tc predicted by $r^2$SCAN, viewed along the c-axis, showing SODTs along [110]. Note that GGA-PBE predicts similar characteristics (not shown).*

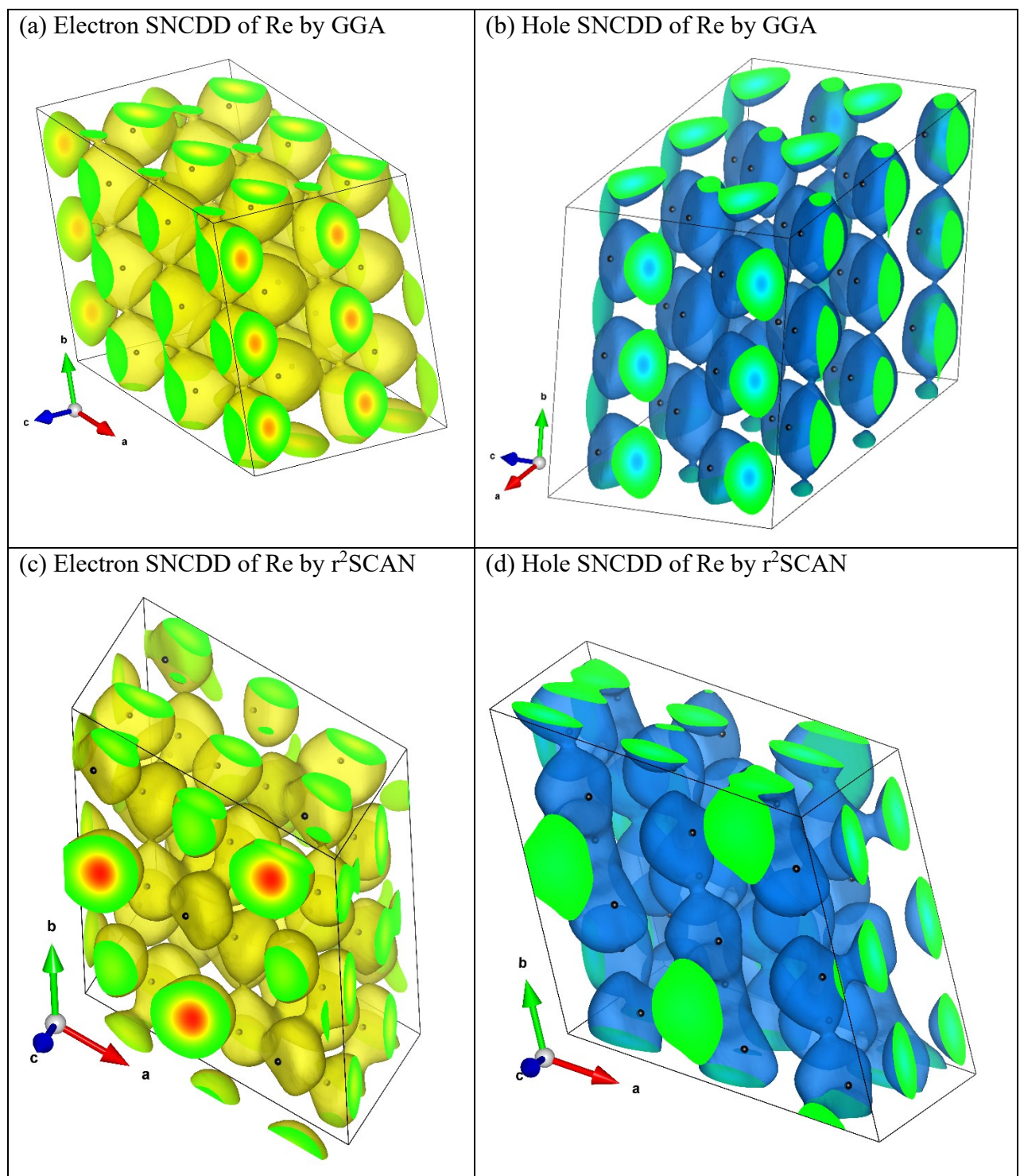

*Figure S 9. Electron (a, yellow) and hole (b, blue) SNCDDs of Re predicted by GGA-PBE, showing SODTs along the b-axis direction. However, $r^2$SCAN does not predict SODTs for Re, see (c) and (d).*

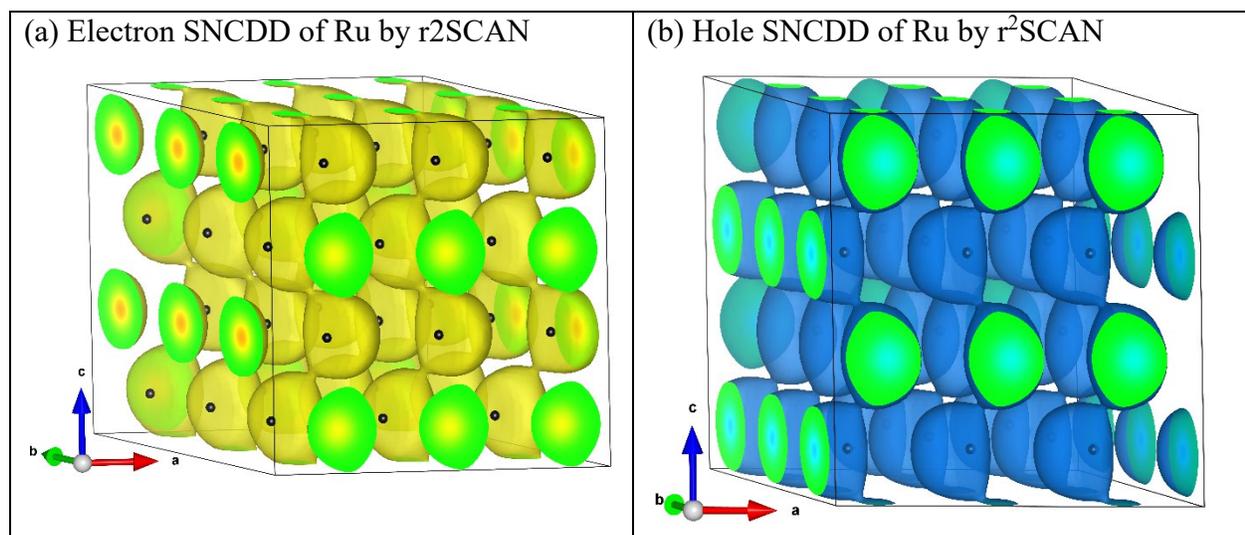

*Figure S 10. Electron (a, yellow) and hole (b, blue) SNCDDs of Ru predicted by r$^2$SCAN, showing SODTs along the c-axis direction with zigzag characters. Note that GGA-PBE predicts similar characteristics (not shown).*

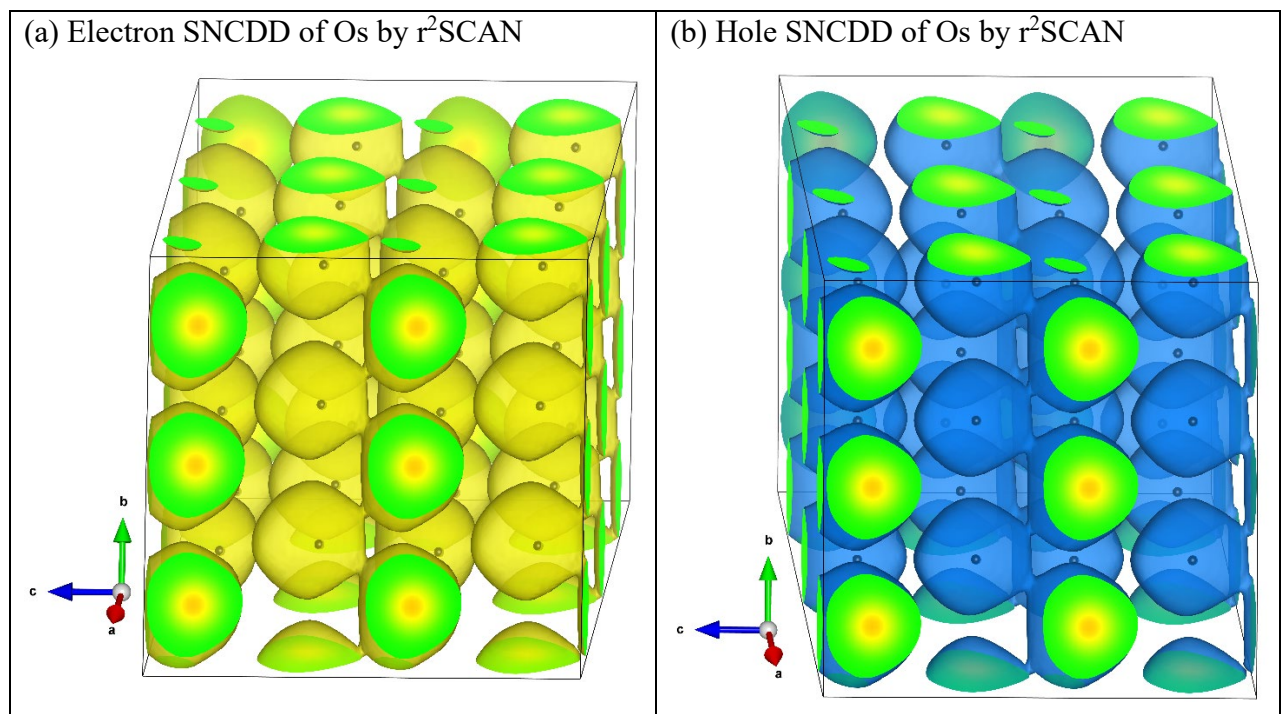

*Figure S 11. Electron (a, yellow) and hole (b, blue) SNCDDs of Os predicted by r²SCAN, showing SODTs along the b-axis. Note that GGA-PBE predicts similar characteristics (not shown).*

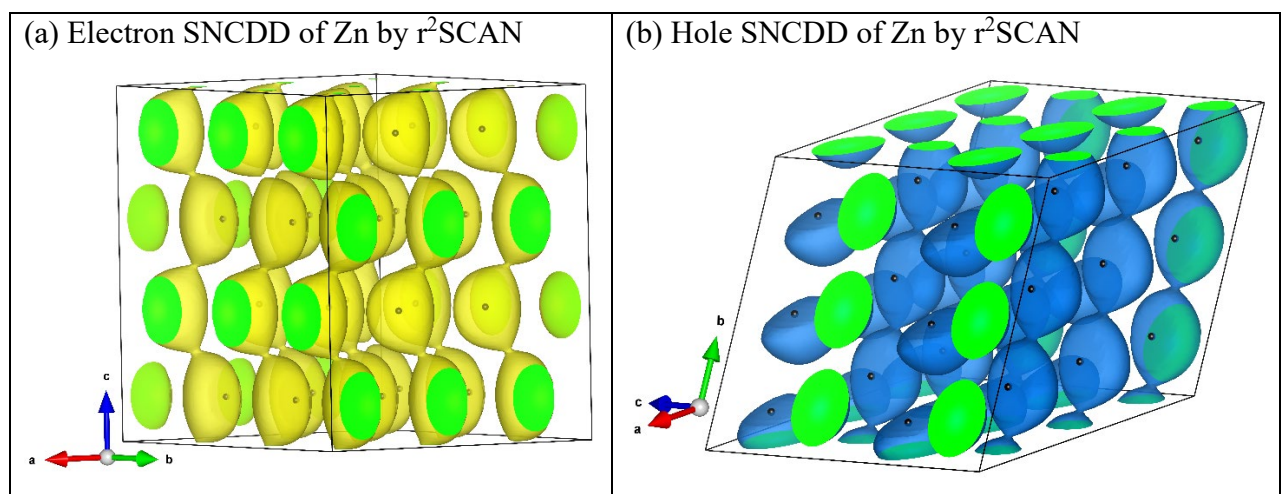

*Figure S 12. Electron (a, yellow) and hole (b, blue) SNCDDs of Zn predicted by r²SCAN, showing SODTs along the b-axis (or the c-axis for electron). Note that GGA-PBE predicts similar characteristics with the formed SODTs along the b-axis (not shown).*

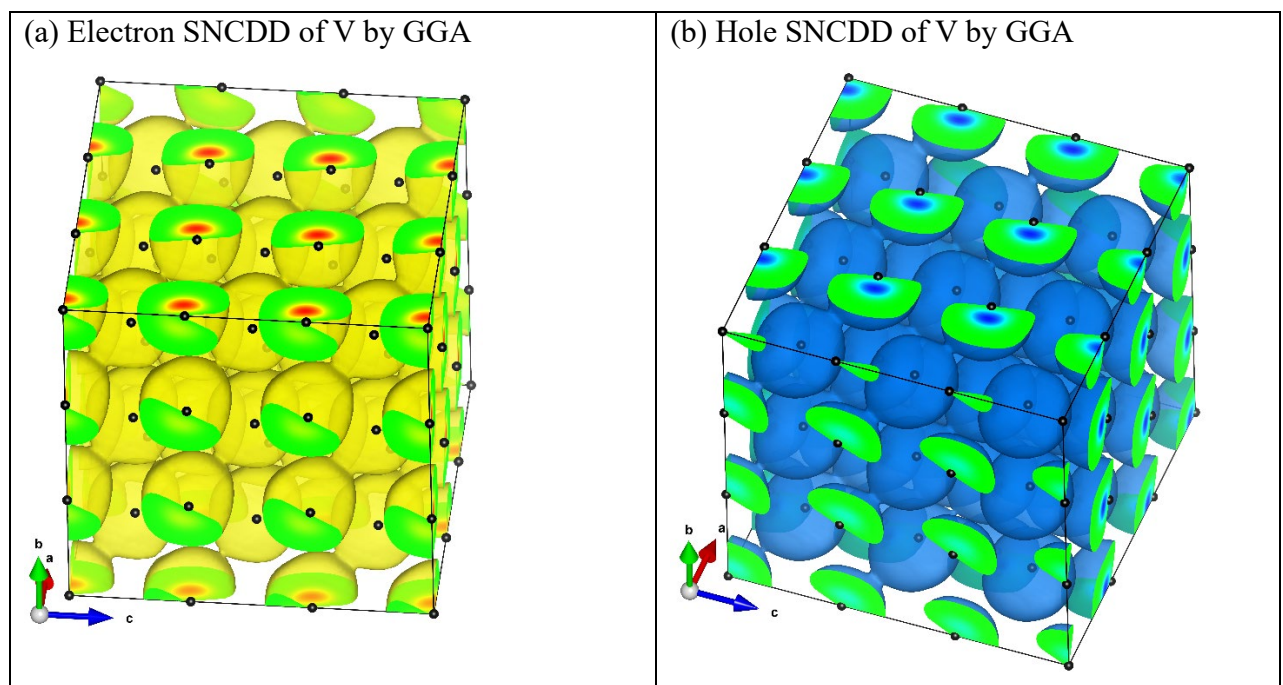

*Figure S 13. Electron (a, yellow) and hole (b, blue) SNCDDs of V predicted by GGA-PBE, showing SODTs along [111]. Note that $r^2$SCAN predictions are shown in Figure 2.*

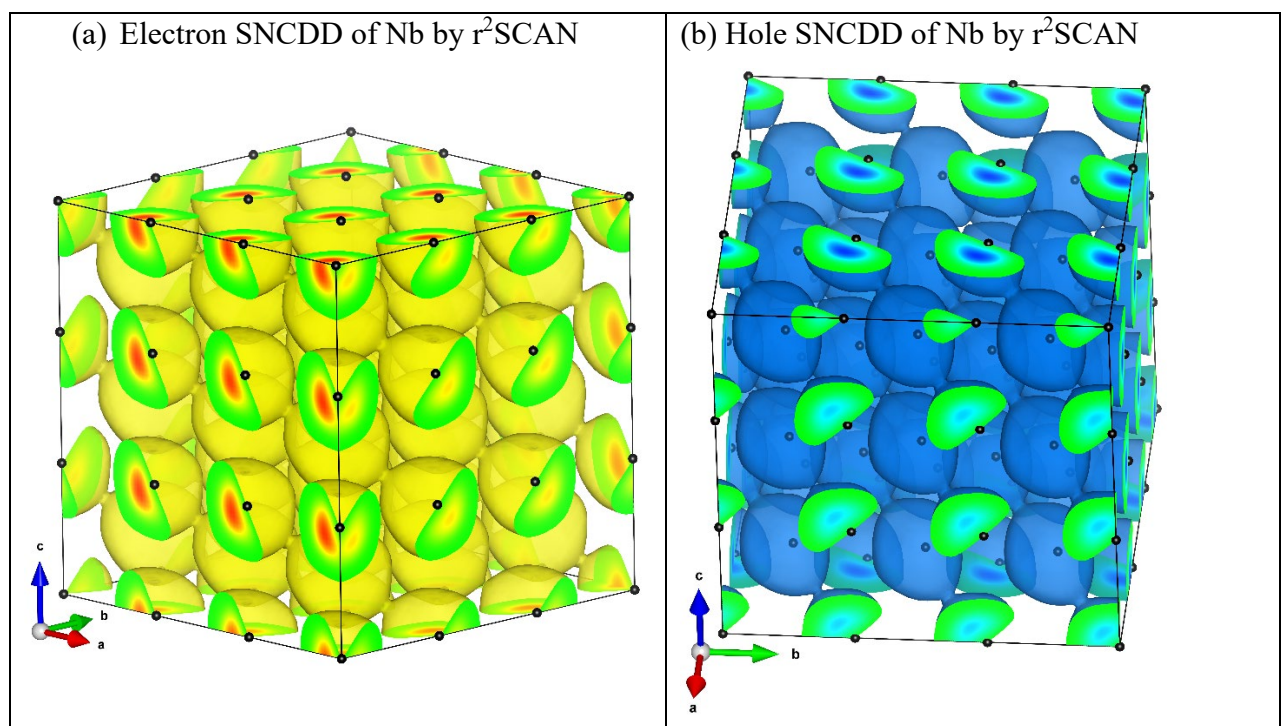

*Figure S 14. Electron (a, yellow) and hole (b, blue) SNCDDs of Nb predicted by $r^2$SCAN, showing SODTs along [$\bar{1}$11]. Note that GGA-PBE predicts similar characteristics (not shown).*

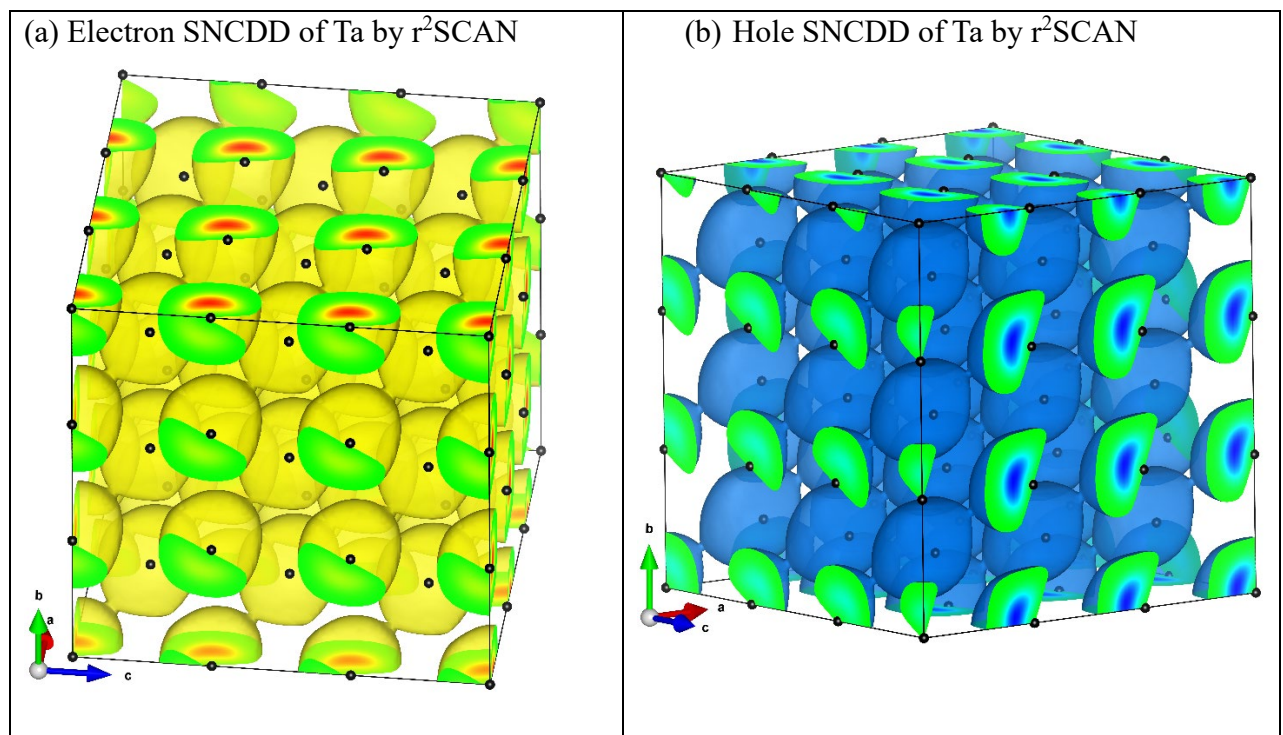

*Figure S 15. Electron (a, yellow) and hole (b, blue) SNCDDs of Ta predicted by r²SCAN, showing SODTs along [111]. Note that GGA-PBE predicts similar characteristics (not shown).*

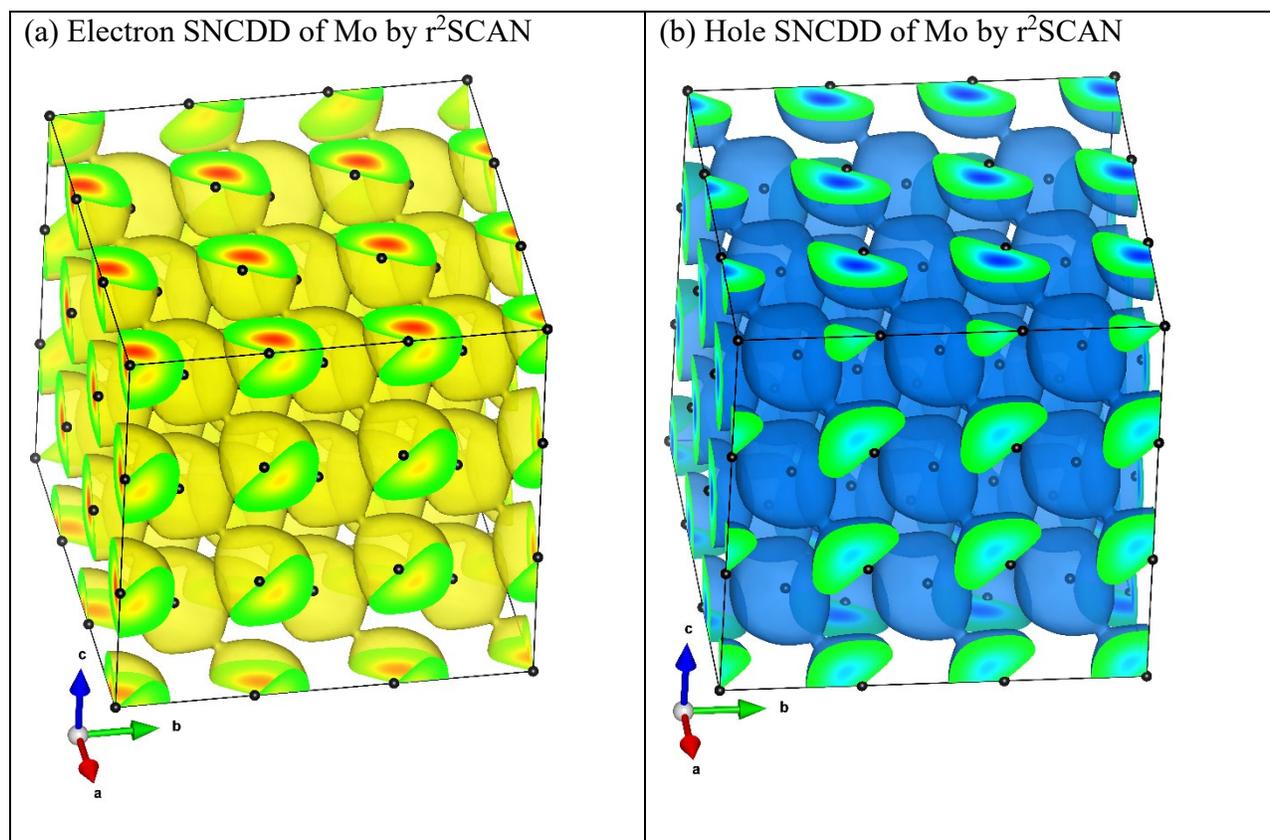

Figure S 16. Electron (a, yellow) and hole (b, blue) SNCDDs of Mo predicted by r²SCAN, showing SODTs along [1̄1̄1]. Note that GGA-PBE predicts roughly similar characteristics but not as clear as the r²SCAN cases.

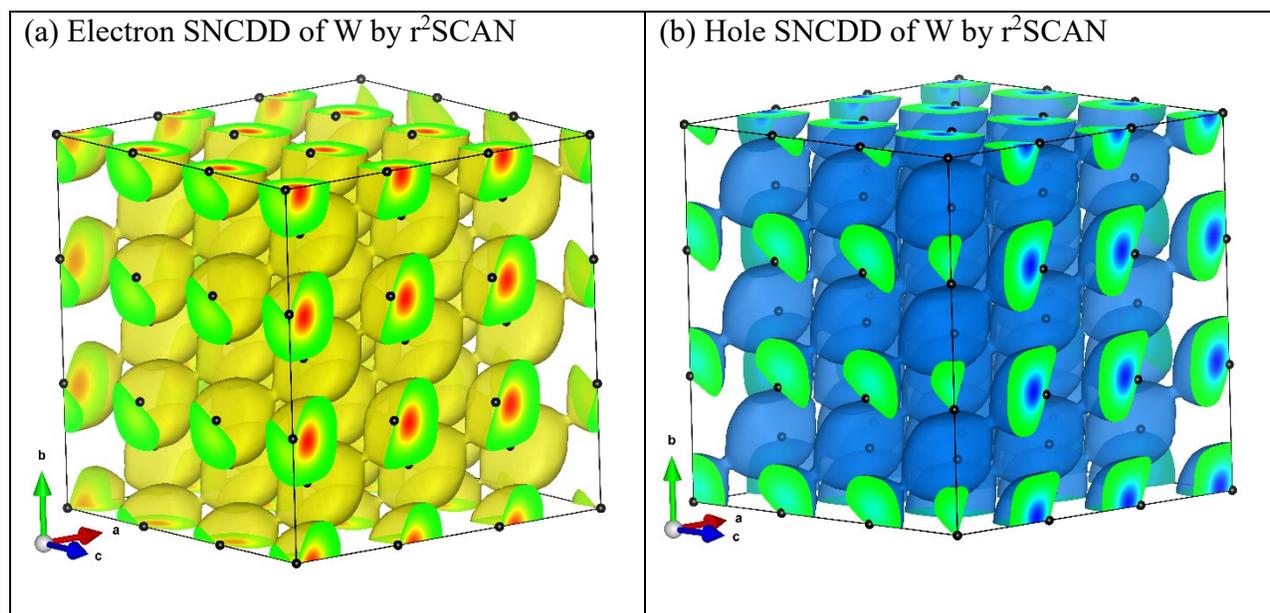

Figure S 17. Electron (a, yellow) and hole (b, blue) SNCDDs of W predicted by $r^2$SCAN, showing SODTs along [111].  Note that GGA-PBE predicts similar characteristics.